\documentclass[11pt, nofootinbib]{article}

\usepackage{amssymb}

\usepackage{amsmath}

\usepackage{bm}

\usepackage{jcappub}

\usepackage{xcolor}

\allowdisplaybreaks

\newcommand{\beq}{\begin{equation}}
\newcommand{\eeq}{\end{equation}}
\newcommand{\bea}{\begin{eqnarray}}
\newcommand{\eea}{\end{eqnarray}}
\newcommand{\ben}{\begin{enumerate}}
\newcommand{\een}{\end{enumerate}}

\newcommand{\pa}{\partial}

\newcommand{\na}{\nabla}
\newcommand{\ed}{{\rm d}}

\newcommand{\ti}{\tilde}
\newcommand{\Rs}{\mathbb{R}}

\renewcommand\({\left(}
\renewcommand\){\right)}
\renewcommand\[{\left[}
\renewcommand\]{\right]}

\newcommand{\ub}{\underbrace}
\newcommand{\os}{\overset}
\newcommand{\us}{\underset}

\newcommand{\nn}{\nonumber}

\newcommand{\bra}{\langle}
\newcommand{\ket}{\rangle}
\newcommand{\Bra}{\left\langle}
\newcommand{\Ket}{\right\rangle}

\newcommand{\al}{\alpha}
\newcommand{\be}{\beta}
\newcommand{\ga}{\gamma}

\newcommand{\de}{\delta}
\newcommand{\De}{\Delta}

\newcommand{\vep}{\varepsilon}

\newcommand{\et}{\eta}
\newcommand{\te}{\theta}
\newcommand{\vte}{\vartheta}

\newcommand{\La}{\Lambda}

\newcommand{\Si}{\Sigma}

\newcommand{\ph}{\phi}
\newcommand{\vph}{\varphi}

\newcommand{\om}{\omega}
\newcommand{\Om}{\Omega}

\newcommand{\Ord}{{\cal O}}

\usepackage{tikz}
\usetikzlibrary{arrows,shapes}
\usetikzlibrary{trees}
\usetikzlibrary{matrix,arrows}
\usetikzlibrary{positioning}
\usetikzlibrary{calc,through}			
\usetikzlibrary{decorations.pathreplacing}  
\usepackage{pgffor}		

\usetikzlibrary{decorations.pathmorphing}
\usetikzlibrary{decorations.markings}
\tikzset{
    lline/.style={solid,line width=1pt,draw=black},
    Lline/.style={solid,line width=0.5pt,draw=black},
    dline/.style={dashed,line width=1pt,draw=black},
    }

\title{General and consistent statistics for cosmological observations}

\author{Ermis Mitsou$^1$,} 
\emailAdd{ermitsou@physik.uzh.ch}
\author{Jaiyul Yoo$^{1,2}$,}
\emailAdd{jyoo@physik.uzh.ch}
\author{Ruth Durrer$^3$,}
\emailAdd{ruth.durrer@unige.ch}
\author{Fulvio Scaccabarozzi$^1$,} 
\author{and Vittorio Tansella$^3$}

\affiliation{$^1$ Center for Theoretical Astrophysics and Cosmology, Institute for Computational Science, University of Zurich, CH--8057 Z\"urich, Switzerland}
\affiliation{$^2$ Physics Institute, University of Z\"urich, Winterthurerstrasse 190, CH--8057, Z\"urich, Switzerland}
\affiliation{$^3$ D\'epartement de Physique Th\'eorique and Center for Astroparticle Physics, Universit\'e de Gen\`eve, 24 quai Ansermet, CH–1211 Gen\`eve 4, Switzerland}

\abstract{This paper focuses on two aspects of the statistics of cosmological observables that are important for the next stages of precision cosmology. First, we note that the theory of reduced angular $N$-point spectra has only been developed in detail up to the trispectrum case and in a fashion that makes it difficult to go beyond. To fill this gap, here we present a constructive approach that provides a systematic description of reduced angular $N$-point spectra and their covariance matrices, for arbitrary $N$. Second, we focus on the common practice in the literature on cosmological observables, which consists in simply discarding a part of the expression, namely, the terms containing fields evaluated at the observer position. We point out that this is not justified beyond linear order in perturbation theory, as these terms contribute to all the multipoles of the corresponding spectra and with a magnitude that is of the same order as the rest of the non-linear corrections. We consider the possibility that the reason for neglecting these terms is a conceptual discomfort when using ensemble averages, which originates in an apparent tension between the ergodic hypothesis and the privileged position of the observer on the light-cone. We clarify this subtle issue by performing a careful derivation of the relation between the theoretical statistical predictions and the observational estimators for all $N$. We conclude that there is no inconsistency whatsoever in ensemble-averaging fields at and near the observer position, thus clearing the way for consistent and robust high-precision calculations.}

\begin{document}

\maketitle

\flushbottom

\section{Introduction}

Upcoming surveys such as \cite{DESI13,LSST04,WFIRST12,SKA09,EUCLID11} will significantly enhance the quantity and quality of the data that we have at our disposal for understanding the universe. This development requires a commensurate effort on the theoretical side for the new observational input to be interpreted correctly. In particular, one needs to consider analytical expressions for cosmological observables beyond the linear order in perturbation theory, as is already clearly reflected by the extensive amount of work on the subject in the case of CMB lensing \cite{Challinor:2005jy,Lewis:2006fu,Hanson:2009kr,Lewis:2011fk,Pettinari:2014iha,Bonvin:2015uha,Marozzi:2016uob,Marozzi:2016und,Marozzi:2016qxl,Pratten:2016dsm,Lewis:2017ans,DiDio:2019rfy}, galaxy number counts \cite{DiDio:2014lka,Bertacca:2014dra,Bertacca:2014wga,DiDio:2015bua,Nielsen:2016ldx,DiDio:2016gpd,Umeh:2016nuh,Jolicoeur:2017nyt,Jolicoeur:2017eyi,Jolicoeur:2018blf,Koyama:2018ttg,DiDio:2018unb,Clarkson:2018dwn,Jalilvand:2019brk,Fuentes:2019nel} and cosmological distances and weak lensing \cite{Bernardeau:2009bm,Bernardeau:2011tc,Umeh:2012pn,Umeh:2014ana,Andrianomena:2014sya,Marozzi:2014kua,Bonvin:2015kea,Fanizza:2018qux,Gressel:2019jxw}. The inclusion of non-linear corrections is not only required for increasing the precision of the basic quantities of interest, such as the power spectrum of observables, but also for obtaining the leading contributions to higher-order statistics (e.g. bispectrum, trispectrum, etc.). The latter contain crucial information about the early universe and the formation of structure, which will become more accessible thanks to the expected advances in observations. In this paper we wish to elaborate on two important aspects of the statistics of cosmological observables for the next stages of precision cosmology which have not yet been considered in the literature. 

The first issue is the lack of generality in our description of angular higher-order statistics. Indeed, in the literature, the rotationally-invariant (or ``reduced") angular $N$-point spectra of a given observable on the sky are defined only up to the trispectrum ($N = 4$) case \cite{HU01a} and built in a rather opaque way that does not generalize easily to arbitrary $N$. The $N = 3,4$ cases are well studied and measured \cite{FESH07,SMSEZA09,SMAMET09,SESMZA10,FECOET15} in CMB observations, whittling down the parameter spaces of the inflationary models. It is known, however, that the hierarchy between $N$-point spectra is not necessarily trivial, e.g. there exist models with a large trispectrum, but negligible bispectrum \cite{Chen:2006nt,Suyama:2007bg,CHHUET09,SUKOFU11}. Therefore, the study of $N$-point spectra beyond $N = 4$ could be relevant for models with both negligible bispectrum and trispectrum. In any case, it is always desirable to have a systematic theoretical framework for approaching a given problem, which is what we will present here. This is a transparent construction of the reduced angular $N$-point spectra, for arbitrary $N$, their covariance matrices for arbitrary $N$ and $N'$, as well as a flurry of useful equations for manipulating them. The central ingredient in this framework will be the generalization of the ``triangular" Wigner $3-j$ symbol to a ``multilateral" one, leading in particular to a neat diagrammatic representation. Our equations will hold for the idealized case of full sky measurements, but they should not be difficult to generalize to the case of partial covering. Note, finally, that the number of components of the angular spectra increases rapidly with $N$, meaning that the new higher-order statistics we can now work with are practically impossible to store, let alone manipulate, in their entirety. Nevertheless, our construction is the indispensable prerequisite for handling these objects, as it provides the starting point for research on privileged subsets, limits, shapes and efficient estimators thereof. 

The second issue we want to discuss concerns the consistency and accuracy of the theoretical predictions, especially at the non-linear orders in perturbation theory which now need to be taken into account. The standard approach to analytical computations of cosmological observables neglects certain terms, namely, those containing fields evaluated at the observer point. As we will discuss, for linear order perturbation theory these observer-dependent contributions have no effect on multipoles $l \geq 2$, but we will argue that they do affect all multipoles at the non-linear level. This is an important point, because none of the works on non-linear effects so far \cite{Challinor:2005jy,Lewis:2006fu,Hanson:2009kr,Lewis:2011fk,Pettinari:2014iha,Bonvin:2015uha,Marozzi:2016uob,Marozzi:2016und,Marozzi:2016qxl,Pratten:2016dsm,Lewis:2017ans,DiDio:2019rfy,DiDio:2014lka,Bertacca:2014dra,Bertacca:2014wga,DiDio:2015bua,Nielsen:2016ldx,DiDio:2016gpd,Umeh:2016nuh,Jolicoeur:2017nyt,Jolicoeur:2017eyi,Jolicoeur:2018blf,Koyama:2018ttg,DiDio:2018unb,Clarkson:2018dwn,Jalilvand:2019brk,Fuentes:2019nel,Bernardeau:2009bm,Bernardeau:2011tc,Umeh:2012pn,Umeh:2014ana,Andrianomena:2014sya,Marozzi:2014kua,Bonvin:2015kea,Fanizza:2018qux,Gressel:2019jxw} considers these observer-dependent terms. Moreover, at the linear level it has already been established that the consideration of the full expressions, i.e. including the observer-dependent terms, has some important advantages, as it leads to results that are gauge-invariant and respect the equivalence principle in $k$-space (absence of spurious infrared divergences) \cite{Yoo:2009au,Yoo:2010ni,Yoo:2014kpa,Yoo:2014sfa,Biern:2016kys,Biern:2017bzo,Yoo:2017svj,Scaccabarozzi:2017ncm,Scaccabarozzi:2018vux,Yoo:2018qba,Grimm:2018nto,Grimm:2020ays}. 

We identify two possible reasons behind the reticence towards considering these observer-dependent terms in the literature. The first one is rooted in a common misconception that observer-dependent terms are synonymous with ``monopole'', because they only affect the later in the power spectrum of the linear theory (or the $s$ multipole for spin $s$ observables). Since low multipoles are dominated by cosmic variance, one is then naturally led to ignore observer-dependent terms. We will show that this is an artifact of linear perturbation theory, because observer-dependent terms actually do arise in non-linear corrections to {\it all} multipoles, and there is no reason to believe that these contributions are negligible. The second possible reason for ignoring these terms might be a discomfort of conceptual nature. Simply put, it is not clear whether one can consistently ensemble average them to compute correlation functions and spectra, because, contrary to the case of source points, one does not probe several ``observer points" observationally for the ergodic hypothesis to apply. A first shaky aspect of this argument is that the ensemble averages that are considered in the standard approach are statistically homogeneous, so this already wipes out the observationally privileged status of the observer point, independently of whether one chooses to include the observer-dependent terms or not. More precisely, we will see that the above objection to the statistical use of observer-dependent terms arises from a misunderstanding of the way the ergodic hypothesis enters the relation between the ensemble averages of the theorist and the geometrical averages of the observer. We will perform a careful derivation of that relation under the standard assumption of statistical homogeneity and isotropy for the stochastic fluctuations, but without any assumptions about which points are included in the ensemble averages. We will then see that the error of this theoretical prediction with respect to the data averages is nothing but cosmic variance. Thus, by carelessly ensemble averaging field products at arbitrary points on the light-cone, independently of how well we can probe them observationally, the theorist is not deviating in any new way from observation other than the fundamental limitation of cosmic variance. Let us also stress that our result is completely general, as we will work with the freshly defined $N$-point spectra for arbitrary $N$. The assumption of full sky measurement will not be problematic here, as the point we wish to make is of qualitative nature. Also, we will only consider the spectra of a single observable, but the formalism generalizes straightforwardly in the case of cross-correlations of different observables.

The paper is organized as follows. In section \ref{sec:preli} we discuss some technical aspects regarding cosmological observables and their theoretical statistical treatment. The reader who is familiar with cosmological observables can skip this part, although some of its definitions and equations will be used later on, so it might be useful to take a quick look anyway. In section \ref{sec:Npoint} we discuss the present situation regarding reduced angular $N$-point spectra in the literature and then derive our formulas for defining and handling these objects for arbitrary $N$. Section \ref{sec:consistency} focuses on the issue of observer-dependent terms. We first discuss in more detail the inconsistencies, and potential inaccuracies, that result from ignoring them in the spectra of observables. Then, using our definition of the angular correlation functions and spectra given in the previous section, we derive the relation between the statistical ones of the theorist and the geometrical ones of the observer. This demonstrates that one is allowed to (and one should) take into account the observer-dependent terms inside ensemble averages. In \ref{sec:extra} we derive the expression for the covariance matrix of two arbitrary reduced angular spectra and determine the asymptotic behavior of cosmic variance for large $l$ values. In \ref{sec:connected} we discuss a few subtleties one should be aware of when working with the connected parts of the statistics. Finally, in section \ref{sec:discussion} we conclude. The appendix contains technical derivations to avoid clogging the main text, as well as the set of Wigner $3-j$ symbol identities we will use in this paper.

\section{Preliminary considerations} \label{sec:preli}

\subsection{Angle, redshift, and observables}

A cosmological observable associated with some localized source of light (galaxy, supernovae, etc.) is a function of two space-time points $\Ord(x_o; x_s)$, the ``observer" point $x_o$ and the ``source" point $x_s$, that are constrained to lie on a common light-like geodesic, with $x_s$ in the past of $x_o$. Thus, for a given observer at $x_o$, the set of possible light source points $x_s$ detected through light forms the past light-cone of $x_o$. The numbers $x_s^{\mu}$ are not observables, as they are ambiguous due to the freedom of performing coordinate transformations. One can of course consider correlation functions of the form $\bra \Ord(x_o; x_1) \dots \Ord(x_o; x_N) \ket$, or the associated spectra in $k$-space, but these cannot be related to observable information, especially in the presence of non-negligible inhomogeneity effects. One must therefore work directly with the physically unambiguous relations, i.e. the relations between observables, so one must parametrize the past-light cone at $x_o$ in terms of observables. One of them is the incoming photon direction in the sky, $n$, leading to an angular parametrization of the sky. Thus, instead of working with the ``$k$-space" spectra, we will work with the ``$l$-space" ones, which are directly related to observable quantities. Finally, for the radial parametrization of the light-cone there are several choices, such as the observed redshift $z$ or the luminosity/angular distances $D_{L,A}$. Here we will consider the former, which is also the most widely used and model independent. Thus, the observer at $x_o$ parametrizes her light-cone in terms of the observables $z$ and $n$, and the physical information lies in the relation between $\Ord$ and $(z,n)$ that is the function $\Ord(x_o;z,n)$. 

The quantities $z$ and $n$ are defined with respect to the observer rest-frame, i.e. a tetrad $e_a$ at $x_o$ 
\beq \label{eq:tetdef}
g_o(e_a, e_b) \equiv \et_{ab} \, ,
\eeq
whose time-component $e_0$ is the observer's 4-velocity, and the source 4-velocity $u_s$ satisfying $g_s(u_s,u_s) \equiv -1$. More precisely, if $k$ denotes the momentum 4-vector of the photon, we have
\beq
1 + z := \frac{g(u_s,k_s)}{g(e_0,k_o)} \, , \hspace{1cm} n_i := -\frac{g(e_i,k_o)}{\sqrt{g(e_j,k_o)\, g(e_j,k_o)}} \, ,   
\eeq
where $i \in \{ 1,2,3 \}$ is the spatial part of the 4-dimensional index $a \in \{ 0,1,2,3 \}$. Note that $n_i$ is the observed angular direction, not the propagation direction, hence the minus sign. These quantities are invariant under space-time coordinate transformations, as any observable should, since a measurement cannot depend on how we parametrize that manifold. In the absence of a tetrad, the only available numbers are the components of $k$ in a coordinate system $k^{\mu}$ and the corresponding angles in the spatial part are the ones an observer uses only if the basis $\pa_{\mu}$ is orthonormal, i.e. only if $g_{\mu\nu}(x_o) = \et_{\mu\nu}$. This is not the case in most of the coordinate systems used in cosmology, which is why one requires a tetrad to obtain the correct angles $n$. Since this field is defined at the observer position, the corresponding correction of the angles it induces will appear as observer terms in the expressions of observables.

The observer tetrad \eqref{eq:tetdef} is defined only up to a Lorentz transformation $e_a \to \La_a^{\,\,\,b} e_b$, but the boosts alter the observer 4-velocity, so the only ambiguity is the orientation of the spatial frame $e_i \to R_i^{\,\,\,j} e_j$, leading to an SO(3) ambiguity for $n$
\beq \label{eq:SO3amb}
n \to R^{-1} n \, .
\eeq 
As for the observable $\Ord$, the theoretical expression must also be a scalar under coordinate transformations \cite{Yoo:2017svj}
\beq \label{eq:Odiff}
\ti{\Ord}(\ti{x}_o;z,n) \equiv \Ord(x_o;z,n) \, ,
\eeq
but it can be a tensor with respect to the Lorentz index $a$ of the observer tetrad. For instance, $\om$ and $n$ are components of the Lorentz vector $k_a := g(e_a,k_o)$ at $x_o$
\beq
(k_a) = -\om \( 1, n_i \) \, ,
\eeq
which is why they transform non-trivially under Lorentz transformations of $e_a$. Since the boost part here is fixed by the definite observer 4-velocity, we only have to deal with the SO(3) ambiguity \eqref{eq:SO3amb}. To that end, we express $n$ in terms of observed angles $(\vte,\vph)$
\beq \label{eq:nofang}
n \equiv \sin \vte \cos \vph\, e_1 + \sin \vte \sin \vph\, e_2 + \cos \vte \, e_3 \, ,
\eeq
so that $\Ord \equiv \Ord(x_o; z,\vte,\vph)$ becomes a function on the unit sphere $S_2$. The latter is a 2-dimensional manifold with coordinates $\vte^A \in \{ \vte, \vph \}$ and with the admissible coordinate transformations  being the ones induced by the rotation \eqref{eq:SO3amb}.\footnote{Note that this manifold is a subset of the tangent space at $x_o$ and with a parametrization that is induced by the tetrad basis $e_a$ of that space, so these angles are not space-time coordinates.} The generic observable will therefore be a tensor field on that manifold $\Ord_{A_1 \dots A_n}$. However, in two dimensions any tensor can be locally reduced to scalars and pseudo-scalars. For instance, a vector can be decomposed into a scalar $v$ and a pseudo-scalar $\ti{v}$ via
\beq
V_A = \na_A v + \vep_A^{\,\,\,B} \na_B \ti{v} \, ,
\eeq      
while a tensor can be decomposed into two scalars $T,t$ and two pseudo-scalars $\ti{T}, \ti{t}$
\beq
T_{AB} = S_{AB} T + \vep_{AB} \ti{T} + \[ \na_A \na_B - \frac{1}{2} \, S_{AB} \na^2 \] t + \vep_{(A}^{\,\,\,\,\,C} \na_{B)} \na_C \ti{t} \, ,
\eeq
where $S_{AB}$, $\vep_{AB}$ and $\na_A$ are the metric, totally anti-symmetric tensor in 2d and covariant derivative on the 2-sphere $S_2$, respectively. Further decomposing these (pseudo-)scalars into spherical harmonics leads to the decomposition of $\Ord_{A_1 \dots A_n}$ into spin-weighted spherical harmonics. To avoid the introduction of the latter, which complicates unnecessarily the formalism, here we will assume that our observables are the (pseudo-) scalars of the above decomposition (up to possible Laplacians). For instance, in the case of the CMB polarization tensor $P_{AB}$ we would directly work with its ``electric" and ``magnetic" components $\na_A \na_B P^{AB}$ and $\vep_A^{\,\,\,C} \na_B \na_C P^{AB}$, respectively.

Finally, another important observable is the incoming photon frequency 
\beq
\om := - g(e_0,k_o) \, ,
\eeq
so the most general parametrization of light-based observables is a spectral distribution $\Ord(x_o; z, n, \om)$. In the case of observables associated with diffuse sources (e.g. the CMB), i.e. that do not have a particular emission moment and therefore no associated redshift, then we simply have $\Ord(x_o; n, \om)$. For the sake of simplicity, here we will assume that $\om$ is either fixed, or that it is integrated over with some given spectral distribution, as in the case of the CMB temperature for instance. We will therefore work with observable functions of the form $\Ord(x_o; z, n)$, but the inclusion of $\om$ is straightforward and basically enters our equations exactly as the redshift dependence. A more detailed account of this subsection's content, and in particular of the underlying geometrical constructions, can be found in \cite{Mitsou:2019nhj}.

\subsection{Ensemble averaged $N$-point correlation functions} \label{sec:ensavN}

The theorist works with tensor fields on the full space-time mani\-fold which we collectively denote by $\Phi(x)$. The observable $\Ord(x_o;z,n)$ is an integro-differential functional of these fields
\beq
\Ord(x_o;z,n) \equiv \Ord(x_o;z,n)[\Phi] \, .
\eeq
Within cosmological perturbation theory, one splits $\Phi$ into a homogeneous and isotropic ``background" solution $\bar{\Phi}$ and a fluctuation $\ph$
\beq
\Phi = \bar{\Phi} + \ph \, ,
\eeq
where the latter is subject to a gauge ambiguity. This is the manifestation of infinitesimal coordinate transformations in this framework, since the background is kept fixed under these transformation. The observable $\Ord(x_o;z,n)$ should, by definition, be independent of the choice of gauge, so one can fix the latter. Moreover, one can use any constraint equations (e.g. the Poisson equation) to reduce the number of fields down to those carrying degrees of freedom, which we denote by $\ph_{\frak{a}}(x)$, i.e. the ``$\frak{a}$" index includes both space-time and internal indices. Thus, the $\ph_{\frak{a}}(t_0,\vec{x})$ data at a given time $t_0$ uniquely determine the ones at any other time $t$
\beq
\ph_{\frak{a}}(t,\vec{x}) \equiv \ph_{\frak{a}}(t,\vec{x})[\ph_{\frak{b}}(t_0,\vec{y})] \, .
\eeq
Here it is assumed that the $\ph_{\frak{a}}$ also contain the momenta/velocities, for fields obeying second-order equations in time. Next, in order to compare with observations, one needs to consider a statistical ensemble of solutions. The $\ph_{\frak{a}}$ are therefore promoted to stochastic fields with a corresponding probability distribution functional (pdf) associating a probability density to each field solution $\ph_{\frak{a}}(x)$. Since the latter are completely determined by their configuration at some reference time $\ph_{\frak{a}}(t_0,\vec{x})$, it suffices to define that pdf on these field configurations $P \equiv P[\ph(t_0)]$ (dropping the index $\frak{a}$ and the position $\vec x$ for notational simplicity). One can then define the moments, i.e. the statistical averages of field products 
\beq \label{eq:Ptmom}
\bra \ph_{\frak{a}_1}(t_0,\vec{x}_1) \dots \ph_{\frak{a}_n}(t_0,\vec{x}_n) \ket := \int D \ph(t_0)\, P[\ph(t_0)]\, \ph_{\frak{a}_1}(t_0,\vec{x}_1) \dots \ph_{\frak{a}_n}(t_0,\vec{x}_n) \, ,
\eeq
which completely determine the functional $P$, and with these one can define the {\it field} correlation functions (FCF) 
\bea
F_{\frak{a}_1 \dots \frak{a}_n} ( x_1, \dots, x_n ) & := & \bra \ph_{\frak{a}_1}(t_1,\vec{x}_1) \dots \ph_{\frak{a}_n}(t_n,\vec{x}_n) \ket \nn \\
 & \equiv & \bra \ph_{\frak{a}_1}(t_1,\vec{x}_1)[\ph(t_0)] \dots \ph_{\frak{a}_n}(t_n,\vec{x}_n)[\ph(t_0)] \ket \, , \label{eq:fieldG}
\eea 
and use the linearity of the averaging operation to express this as a functional of the \eqref{eq:Ptmom}. In particular, $P$ is chosen such that
\beq
\bra \ph_{\frak{a}}(x) \ket \equiv 0 \, ,
\eeq
which can be alternatively stated as
\beq
\bra \Phi(x) \ket \equiv \bar{\Phi}(x) \, .
\eeq
With the above definitions one can now perform statistical averages of arbitrary functionals of the $\ph_{\frak{a}}(x)$. In particular, the {\it theoretical} correlation functions (TCF) of the observables are defined by 
\beq  \label{eq:Gthdef}
G^{\rm th} \( x_o; \{ z_k, n_k \}_{k=1}^N \) := \Bra \prod_{k=1}^N \Ord(x_o; z_k, n_k)[\Phi] \Ket \, ,
\eeq
which are therefore ultimately a functional of  $\bar{\Phi}$ and the FCFs \eqref{eq:fieldG}. Now note that, although strict homogeneity and isotropy are lost as soon as $\ph \neq 0$, these notions can be reintroduced at the statistical level by imposing the corresponding symmetries on $P[\ph(t_0)]$. Thus, we require that, if the two configurations $\ph_{\frak{a}}(t_0)$ and $\ph'_{\frak{a}}(t_0)$ are related by an isometry of the background geometry, then $P[\ph(t_0)] = P[\ph'(t_0)]$. As a result, the FCFs and TCFs are invariant under isometries, meaning in particular that $G^{\rm th}$ is independent of the observer position $\vec{x}_o$
\beq \label{eq:Gthtrans}
G^{\rm th} \( x_o; \{ z_k, n_k \}_{k=1}^N \) \equiv G^{\rm th} \( t_o; \{ z_k, n_k \}_{k=1}^N \) \, ,
\eeq
and invariant under a common rotation of its $N$ directions
\beq \label{eq:Gthiso}
G^{\rm th} \( t_o; \{ z_k, R n_k \}_{k=1}^N \) \equiv G^{\rm th} \( t_o; \{ z_k, n_k \}_{k=1}^N \) \, .
\eeq

\section{The general theory of reduced angular $N$-point spectra} \label{sec:Npoint}

\subsection{The previous implicit approach} \label{sec:statusquo}

Consider some observable on the sky $\Ord(n)$, where for the purposes of the present discussion we can omit the redshift dependence. The quantity that is usually considered in theory is the angular $N$-point spectrum through ensemble averaging $\Bra \Ord_{l_1m_1} \dots \Ord_{l_N m_N} \Ket$, where $\Ord_{lm}$ are the spherical harmonic components. In the standard approach where statistical isotropy is assumed, that $N$-point spectrum is invariant under rotations, so that not all of its components are independent. Its information can therefore be stored in a rotationally-invariant quantity with fewer indices called the ``reduced" angular spectrum. Assuming for simplicity here $\Bra \Ord_{lm} \Ket = 0$, the lowest-order cases are the angular power spectrum
\beq \label{eq:poex}
\left\langle\Ord_{l_1m_1}\Ord_{l_2m_2} \right\rangle = \de_{l_1 l_2} \de_{m_1,-m_2} (-1)^{m_2}\, C_{l_1} \, ,
\eeq
where $C_l$ is the ``reduced angular power spectrum", and the angular bispectrum
\beq \label{eq:bisex}
\left\langle\Ord_{l_1m_1}\Ord_{l_2m_2}\Ord_{l_3m_3}\right\rangle
= \left(\begin{array}{ccc}l_1&l_2&l_3\\ m_1&m_2&m_3\end{array}\right)
B_{l_1l_2l_3} \, ,
\eeq
where $B_{l_1l_2l_3}$ is the ``reduced angular bispectrum" and the first factor is the Wigner $3-j$ symbol. As one could expect from isotropy, $C_l$ and $B_{l_1l_2l_3}$ have no $m_k$ dependence. The relations \eqref{eq:poex} and \eqref{eq:bisex} can be inverted and, in particular, allow one to build unbiased estimators for these quantities
\beq \label{eq:estimex}
\hat{C}_l := \frac{1}{2l+1}\, \sum_m \Ord_{lm} \Ord^*_{lm} \, , \hspace{1cm} \hat{B}_{l_1l_2l_3} = \sum_{m_1,m_2,m_3} \left(\begin{array}{ccc}l_1&l_2&l_3\\ m_1&m_2&m_3\end{array}\right) \Ord_{l_1m_1}\Ord_{l_2m_2}\Ord_{l_3m_3} \, ,
\eeq 
i.e.
\beq \label{eq:thofobs}
\bra \hat{C}_l \ket = C_l \, , \hspace{1cm} \bra \hat{B}_{l_1l_2l_3} \ket = B_{l_1l_2l_3} \, .
\eeq
One can therefore work exclusively with the invariant and more compact reduced spectra, both at the theoretical and data level, instead of the full ones. The theory of the reduced angular $N$-point spectra is well developed in the literature up to the trispectrum case \cite{OKHU02,FESH07,SMSEZA09,SMAMET09,SESMZA10,RESHFE10,KASMHE11,SMKA12, FECOET15}, based on the pioneering work in \cite{HU01a} which considered the following method. First, one looks for the relation between the full spectrum and its reduced counterpart (e.g. \eqref{eq:poex} and \eqref{eq:bisex}) by imposing on the former the invariance under rotations
\beq \label{eq:exeqbis}
\left\langle\Ord_{l_1m_1} \dots \Ord_{l_N m_N}\right\rangle
=\sum_{m_k'}\left\langle\Ord_{l_1m'_1} \dots \Ord_{l_N m'_N}
\right\rangle\times D_{l_1,m'_1m_1} \dots D_{l_N,m'_Nm_N} \, ,
\eeq
where $D_{l,mm'}$ is the Wigner matrix, all of them here depending on the same rotation angles. Using the identities of the Wigner matrices and $3-j$ symbols, one must then find the most general solution to this equation, which is \eqref{eq:poex} and \eqref{eq:bisex} for the $N = 2,3$ cases, respectively. Finally, one must also invert this relation to obtain its estimator (e.g. \eqref{eq:estimex}) and, through that quantity, its explicit definition in terms of ensemble averages of the observables (e.g. \eqref{eq:thofobs}). Although this procedure generalizes straightforwardly to $N = 4$ \cite{HU01a} and beyond\footnote{Following this procedure, in \cite{ABPE10} the authors have found the analogue of Eqs. \eqref{eq:poex} and \eqref{eq:bisex} for the quadrispectrum case $N = 5$. In \cite{LACAS14} one can find a formal expression for arbitrary $N$, and also the real space $N$-point correlation function in terms of the reduced angular spectrum, but without rigorous demonstration. In both of these references, however, the inversion to find the estimators and ensemble-averaged spectra in terms of the observables, i.e. the analogues of \eqref{eq:estimex} and \eqref{eq:thofobs}, is not given.}, it becomes quickly very difficult in practice, as one has to ``guess" the relation between the full and reduced spectrum by inferring it from an increasingly complicated equation. Moreover, even after finding a solution, one must still prove that it is the most general one and then invert it, which is also harder to do for larger values of $N$. As a result, the generalization of \eqref{eq:estimex} for arbitrary $N$ does not exist in the literature. Finally, with this procedure the reduced spectrum is determined only up to an overall normalization, which can depend on the $l_i$ numbers, since only the $m_i$ numbers are summed over in the defining equation \eqref{eq:exeqbis}.\footnote{See for instance \cite{LACAS14,RESHFE10}, where alternative normalizations are considered and are essentially related to the $l_i$-dependent factors between the Gaunt and $3-j$ coefficients.}

\subsection{A new constructive approach} \label{sec:constr}

Here we propose a computationally straightforward approach to the problem. Since the defining property behind the notion of reduced spectra is rotational invariance, we will first focus on this geometrical aspect, i.e. before any statistical considerations. We will therefore be constructing the generalization of the full estimators \eqref{eq:estimex}, which exclusively depend on observed data, with their statistical counterparts being obtainable by simply taking the ensemble average as in \eqref{eq:thofobs}. In fact, we will refer to these estimators as the ``observational" reduced spectra, as opposed to the ``theoretical" ones \eqref{eq:Gthdef} which are obtained though statistical ensembles that are observationally unavailable. This is to highlight the fact that the observational reduced spectra have a significance of their own, as they amount to all the rotationally-invariant information one can extract out of the observed data. This is the physical information, as it is independent of the artificial SO(3) ambiguity of the observer's spatial reference frame $e_i$ in \eqref{eq:nofang}.  

We start with the fact that the observer only has access to a single light-cone $x_o$ and therefore to the observables $\Ord(x_o;z,n)$. The building blocks for rotationally-invariant functionals of $\Ord(x_o;z,n)$ are the average of the products $\prod_{k=1}^N \Ord(x_o; z_k, n_k)$ over all possible common rotations of the directions $n_k$, i.e. the average over the SO(3) group
\beq \label{eq:Gobformal}
\frac{\int \ed R \, \prod_{k=1}^N \Ord(x_o; z_k, R^{-1} n_k)}{\int \ed R} \, , 
\eeq
where $\ed R$ is the Haar measure on SO(3) and is invariant under group multiplication. Thanks to this, if one rotates the $n_k$ in \eqref{eq:Gobformal} with the same rotation $R'$, then this can be reabsorbed in the dummy variable $R$ by the redefinition $R \to R' R$, thus leaving the average invariant. To turn \eqref{eq:Gobformal} into a well-defined integral, we consider an arbitrary orthonormal reference frame $e_i$ and parametrize $R$ as
\beq
R ( \al, \be, \ga ) := R_3 ( \al ) \, R_2 ( \be ) \, R_3 ( \ga ) \, ,
\eeq
where $R_i(\te)$ denotes the matrix corresponding to a rotation around $e_i$ with angle $\te$, so that $\al,\be,\ga$ are the Euler angles  
\beq \label{eq:Euler}
\al \in [\,0,2\pi\,[ \, \, , \hspace{1cm} \be \in [\,0,\pi\,] \, \, , \hspace{1cm} \ga \in [\,0,2\pi\,[ \, \, .
\eeq
In particular, the inverse matrix is simply 
\beq \label{eq:Rinv}
R^{-1} ( \al, \be, \ga ) \equiv R ( -\ga, -\be, -\al )  \, .
\eeq
The Haar measure $\ed R$ on SO(3) now reads
\beq \label{eq:HaarSO3}
\ed R(\al,\be,\ga) \equiv \sin \be\, \ed \al\, \ed \be\, \ed \ga \, ,
\eeq
so the average over SO(3) of some function $f(n_1, \dots, n_N)$ is given by
\bea \label{eq:SO3avdef}
\Bra f(n_1, \dots, n_N) \Ket_{\rm SO(3)} &:=&  \\
&& \hspace*{-3cm} \frac{1}{8\pi^2} \int_0^{2\pi} \ed \al \int_0^{\pi} \sin \be\, \ed \be \int_0^{2\pi} \ed \ga \, f\( R^{-1}(\al,\be,\ga)\,n_1, \dots, R^{-1}(\al,\be,\ga)\,n_N \) \, , \nn
\eea
and thus, the $N$-point
{\it observational} correlation functions (OCF)
are defined by
\beq \label{eq:Gobdef}
G^{\rm ob} \( x_o; \{ z_k, n_k \}_{k=1}^N \) := \Bra \prod_{k=1}^N \Ord(x_o; z_k, n_k) \Ket_{\rm SO(3)} \, . 
\eeq
In particular, the $N=1$ case reduces to the average over the sphere
\beq
G^{\rm ob}(x_o; z) \equiv \frac{1}{4\pi} \int \ed \Om\, \Ord\( x_o; z, n \) \, ,
\eeq
which is shown by picking $e_3 = n$. For $N > 1$, three out of the $2N$ angles in $G^{\rm ob}$ are redundant, since we are free to rotate at will, or equivalently, to choose the reference frame $\{ e_i \}_{i=1}^3$ arbitrarily. For instance, one can pick (assuming that $n_1$ and $n_2$ are not parallel)
\beq
e_3 = n_1 \, , \hspace{1cm} e_2 = \frac{n_2 - (n_1 \cdot n_2)\, n_1}{\sqrt{1 - (n_1 \cdot n_2)^2}} \, , \hspace{1cm} e_1 = \frac{n_1 \times n_2}{\sqrt{1 - (n_1 \cdot n_2)^2}} \, ,
\eeq
thus leaving us with a dependence on the $2N-3$ angles $\vte_2, \dots \vte_N$ and $\vph_3, \dots, \vph_N$ that parametrize $\{ n_k \}_{k=2}^N$ in the $e_i$ basis \eqref{eq:nofang}.

Now note that the $N$-point OCF is a redshift-dependent function on
\beq
S_2^N := \us{N\,\,{\rm times}}{\ub{S_2 \times \dots \times S_2}} \, ,
\eeq
that is symmetric under a common rotation of these $N$ spheres, just as its theoretical counterpart \eqref{eq:Gthiso}. In particular, expressing $\Ord$ inside $G^{\rm ob}$ in terms of the stochastic fields $\Ord \equiv \Ord[\Phi]$ and using the linearity of averaging, we obtain the identity
\beq \label{eq:ensofGob}
\Bra G^{\rm ob} \( x_o; \{ z_k, n_k \}_{k=1}^N \) \Ket \equiv G^{\rm th}\( t_o; \{ z_k, n_k \}_{k=1}^N \) \, , 
\eeq
showing that the $G^{\rm ob}$ are full unbiased estimators of the $G^{\rm th}$. The idea now is to decompose such functions in a basis of SO(3)-invariant functions on $S_2^N$. For $N=2$ the only SO(3)-invariant combination of the two vectors is $n_1 \cdot n_2 \in [-1,1]$ and a basis of functions on that interval are the Legendre polynomials $P_l$, so
\beq \label{eq:GobofGobl}
G^{\rm ob} \( x_o; z_1, z_2, n_1, n_2 \) \equiv G^{\rm ob}_l \( x_o; z_1, z_2 \) \( 2 l +1 \) P_l \( n_1 \cdot n_2 \) \, , 
\eeq
where 
\beq \label{eq:Gob2l}
G^{\rm ob}_l \( x_o; z_1, z_2 \) :=  \frac{\Ord_{lm}(x_o; z_1)\, \Ord^*_{lm}(x_o; z_2)}{2l + 1} \, ,
\eeq
is the observational 2-point spectrum and $\Ord_{lm}$ are the harmonic components of $\Ord$
\beq \label{eq:fdecomp}
\Ord(x_o; z, n) \equiv \Ord_{lm}(x_o; z)\, Y_{lm}(n) \, ,
\eeq
Note that we work with the unit-average normalization of the spherical harmonics \eqref{eq:Ynorm}, which is the natural one in this context, and that we keep the summation over $l,m$ indices implicit for notational simplicity. In the presence of both dummy and free $l,m$ indices, their nature will be inferable unambiguously by looking at both sides of the equation. The $m$ indices will always be clearly associated to some $l$ value and therefore run from $-l$ to $l$, while the $l$ indices run from $s$ to $\infty$, where $s$ is the spin of the observable under consideration.\footnote{For instance, we have $s = 0$ for CMB temperature maps, while $s=2$ for the maps of the electric and magnetic components of the polarization field.} For $N > 2$, the decomposition is performed in detail in appendix \ref{sec:hdec} for generic functions $f(n_1, \dots, n_N)$ and the result is the following
\beq \label{eq:GobofGoblN}
G^{\rm ob} \( x_o; \{ z_k, n_k \}_{k=1}^N \) \equiv G^{\rm ob}_{l_1 \dots l_N| L_1 \dots L_{N-3}} \( x_o; \{ z_k \}_{k=1}^N \) Y_{l_1 \dots l_N| L_1 \dots L_{N-3}} \( \{ n_k \}_{k=1}^N \)  \, , 
\eeq
where
\beq  \label{eq:GobNl}
G^{\rm ob}_{l_1 \dots l_N| L_1 \dots L_{N-3}} \( x_o; \{ z_k \}_{k=1}^N \) := W^{l_1 \dots l_N| L_1 \dots L_{N-3}}_{m_1 \dots m_N} \prod_{k=1}^N \Ord_{l_k m_k}(x_o; z_k) \, ,
\eeq
and
\beq \label{eq:GenYbasis}
Y_{l_1 \dots l_N| L_1 \dots L_{N-3}} \( \{ n_k \}_{k=1}^N \) := W^{l_1 \dots l_N| L_1 \dots L_{N-3}}_{m_1 \dots m_N} \prod_{k=1}^N Y_{l_k m_k}(n_k)  \, . 
\eeq
and where we have defined the following coefficients
\bea
W^{l_1 \dots l_N|L_1 \dots L_{N-3}}_{m_1 \dots m_N} & := & \( \begin{array}{ccc} l_1 & l_2 & L_1 \\ m_1 & m_2 & -M_1 \end{array} \) \[ \prod_{k=1}^{N-4} (-1)^{L_k+M_k} \sqrt{2L_k+1} \( \begin{array}{ccc} L_k & l_{k+2} & L_{k+1} \\ M_k & m_{k+2} & -M_{k+1} \end{array} \) \] \nn \\
 & & \times (-1)^{L_{N-3} + M_{N-3}} \sqrt{2L_{N-3}+1} \( \begin{array}{ccc} L_{N-3} & l_{N-1} & l_N \\ M_{N-3} & m_{N-1} & m_N \end{array} \) \, . \label{eq:Wdef}
\eea
Although the latter appear as a sum over the $M_k$ indices, they are actually a single product because a Wigner $3-j$ symbol vanishes if the sum of its $m_k$ entries is not zero. Also for that reason, on can directly check that these coefficients are non-zero only if $\sum_{k=1}^N m_k = 0$. 

Equations \eqref{eq:GobofGoblN} to \eqref{eq:Wdef} are the main result of this section. Note that they can actually also encompass the $N = 1,2$ cases if we define the redundant notation
\bea
G^{\rm ob}_{l_1} \( x_o; z_1 \) & := & \de^0_{l_1} G^{\rm ob} \( x_o; z_1 \) \, , \nn \\
G^{\rm ob}_{l_1l_2} \( x_o; z_1, z_2 \) & := & (-1)^{l_1} \de_{l_1 l_2} \sqrt{2l_1+1}\,G_{l_1}^{\rm ob} \( x_o; z_1, z_2 \) \, , \label{eq:Gob2spec}
\eea
\bea
Y_{l_1}(n_1) & := & \de^0_{l_1} \, , \nn \\
Y_{l_1 l_2}(n_1, n_2) & := & (-1)^{l_1} \de_{l_1 l_2} \sqrt{2 l_1 + 1}\, P_{l_1}(n_1 \cdot n_2) \, ,  \label{eq:Yll2spec}
\eea
and
\bea 
W^{l_1}_{m_1} & := & \( \begin{array}{ccc} l_1 & 0 & 0 \\ m_1 & 0 & 0 \end{array} \) \equiv \de^{l_1}_0 \de^0_{m_1} \, , \nn \\
W^{l_1 l_2}_{m_1 m_2} & := & \( \begin{array}{ccc} l_1 & l_2 & 0 \\ m_1 & m_2 & 0 \end{array} \) \equiv \frac{(-1)^{l_1+m_1}}{\sqrt{2 l_1 + 1}}\, \de^{l_1l_2} \de_{m_1, -m_2} \, .  \label{eq:Wdef12}
\eea
The ``harmonic" components of the $N$-point OCF \eqref{eq:GobNl} are the observational $N$-point spectra which generalize the structures \eqref{eq:estimex} to arbitrary $N$. In particular, for the $N = 3,4$ cases we recover the known results \cite{HU01a}, up to a different normalization convention factor $(-1)^L \sqrt{2L + 1}$ for $N = 4$
\bea
G^{\rm ob}_{l_1 l_2 l_3} \( x_o; z_1, z_2, z_3 \) & \equiv & \( \begin{array}{ccc} l_1 & l_2 & l_3 \\ m_1 & m_2 & m_3 \end{array} \) \prod_{k=1}^3 \Ord_{l_k m_k}(x_o; z_k) \, ,  \\
G^{\rm ob}_{l_1 l_2 l_3 l_4 | L} \( x_o; z_1, z_2, z_3, z_4 \) & \equiv & (-1)^{L+M} \sqrt{2L+1} \( \begin{array}{ccc} l_1 & l_2 & L \\ m_1 & m_2 & -M \end{array} \) \( \begin{array}{ccc} L & l_3 & l_4 \\ M & m_3 & m_4 \end{array} \) \nn \\
 & & \times \prod_{k=1}^4 \Ord_{l_k m_k}(x_o; z_k) \, . 
\eea
As for the functions \eqref{eq:GenYbasis}, they form indeed a basis for SO(3)-invariant functions on $S_2^N$, as shown in appendix \ref{sec:hdec}. They are orthonormal \eqref{eq:newYbortho} and therefore we can invert \eqref{eq:GobofGoblN} 
\bea 
G^{\rm ob}_{l_1 \dots l_N| L_1 \dots L_{N-3}} \( x_o; \{ z_k \}_{k=1}^N \) & \equiv & \int \(  \prod_{k=1}^N \frac{\ed \Om_k}{4\pi} \) Y^*_{l_1 \dots l_N| L_1 \dots L_{N-3}}\( \{ n_k \}_{k=1}^N \) \nn \\
 & & \hspace{2.6cm} \times G^{\rm ob} \( x_o; \{ z_k, n_k \}_{k=1}^N \) \, . \label{eq:GoblofGobN}
\eea
Observe how the quantities \eqref{eq:GobNl} and \eqref{eq:GenYbasis} are explicitly SO(3)-invariant, since they only depend on total angular momentum numbers $l_k$ and $L_k$, making a total of $2N-3$ indices. As noticed earlier, this is indeed the number of independent angles present in the corresponding correlation functions. Note also that \eqref{eq:GobNl} and \eqref{eq:GenYbasis} are the very same combinations of the observable components and spherical harmonics, respectively. As we will see later on, all of the non-trivial contractions of indices will always be controlled by the coefficients \eqref{eq:Wdef} with no extra factors, which comes from the fact that they are orthonormal in the sense of \eqref{eq:Wortho}. Thus, contrary to the implicit method described in the previous subsection \ref{sec:statusquo}, here there exists a normalization of the reduced spectra that appears as naturally privileged for computational convenience. 

For the sake of completeness, let us also present the generalization of \eqref{eq:thofobs} by reminding that the TCFs \eqref{eq:Gthdef} are SO(3)-invariant functions on $S_2^N$ (see \eqref{eq:Gthiso}), so they can also be decomposed in the basis \eqref{eq:GenYbasis}. Using \eqref{eq:ensofGob} and the linearity of averaging we find 
\bea
G^{\rm th}_{l_1 \dots l_N| L_1 \dots L_{N-3}} \( t_o; \{ z_k \}_{k=1}^N \) & := & \int \(  \prod_{k=1}^N \frac{\ed \Om_k}{4\pi} \) Y^*_{l_1 \dots l_N| L_1 \dots L_{N-3}}\( \{ n_k \}_{k=1}^N \) G^{\rm th} \( t_o; \{ z_k, n_k \}_{k=1}^N \) \nn \\
 & \equiv & \Bra G^{\rm ob}_{l_1 \dots l_N| L_1 \dots L_{N-3}} \( x_o; \{ z_k \}_{k=1}^N \) \Ket \nn \\
 & \equiv & W^{l_1 \dots l_N| L_1 \dots L_{N-3}}_{m_1 \dots m_N}  \Bra \prod_{k=1}^N \Ord_{l_k m_k}(x_o; z_k) \Ket \, . \label{eq:GobensGth}
\eea
Finally, to make contact with the implicit method described in subsection \ref{sec:statusquo}, let us also invert equation \eqref{eq:GobensGth} to obtain the generalization of \eqref{eq:poex} and \eqref{eq:bisex} 
\bea
\Bra \prod_{k=1}^N \Ord_{l_k m_k}(z_k) \Ket & \equiv & \int \[ \prod_{k=1}^N \frac{\ed \Om_k}{4\pi}\, Y^*_{l_k m_k}(n_k) \] \Bra \prod_{k=1}^N \Ord(z_k,n_k) \Ket \nn \\
 & \equiv & \int \[ \prod_{k=1}^N \frac{\ed \Om_k}{4\pi}\, Y^*_{l_k m_k}(n_k) \] G^{\rm th} \( \{ z_k, n_k \}_{k=1}^N \) \nn \\
 & \equiv & G^{\rm th}_{l_1 \dots l_N| L_1 \dots L_{N-3}} \( \{ z_k \}_{k=1}^N \) \nn \\
 & & \times \int \[ \prod_{k=1}^N \frac{\ed \Om_k}{4\pi}\, Y^*_{l_k m_k}(n_k) \]  Y_{l_1 \dots l_N| L_1 \dots L_{N-3}} \( \{ n_k \}_{k=1}^N \) \nn \\
 & \equiv & W^{l_1 \dots l_N| L_1 \dots L_{N-3}}_{m'_1 \dots m'_N} G^{\rm th}_{l_1 \dots l_N| L_1 \dots L_{N-3}}\( \{ z_k \}_{k=1}^N \) \nn \\
 & & \times \int \[ \prod_{k=1}^N \frac{\ed \Om_k}{4\pi}\, Y^*_{l_k m_k}(n_k)\, Y_{l_k m'_k}(n_k) \]  \nn \\ 
 & \equiv & W^{l_1 \dots l_N| L_1 \dots L_{N-3}}_{m_1 \dots m_N} G^{\rm th}_{l_1 \dots l_N| L_1 \dots L_{N-3}}\( \{ z_k \}_{k=1}^N \) \, . \label{eq:genfulltored}
\eea
This is the general solution to the equation of the implicit method \eqref{eq:exeqbis} for arbitrary $N$, up to a $l_k, L_k$-dependent normalization. This result highlights again the naturalness of the normalization used here for the reduced angular spectra, as all the complicated factors entering any of the above equations always combine to form the multilateral Wigner symbols exactly. Note also that, with our construction, the consideration of the full spectra \eqref{eq:genfulltored} becomes superfluous, as one no longer requires them to derive the reduced ones as in the implicit method of the previous subsection.

\subsection{Multilateral diagrammatic representation}

\begin{figure}[h!] 
\begin{center}
\begin{tikzpicture}[scale = 2]
 \begin{scope}
\draw[lline] (0,0) -- (1,1.5);
\draw[lline] (1,1.5) -- (3,2);
\draw[lline] (3,2) -- (4.5,0.5);
\draw[lline] (4.5,0.5) -- (3,-0.6);
\draw[lline] (3,-0.6) -- (4,-2);
\draw[lline] (4,-2) -- (1,-1.5);
\draw[lline] (1,-1.5) -- (0,0);
\draw[Lline] (0,0) -- (3,2);
\draw[Lline] (0,0) -- (4.5,0.5);
\draw[Lline] (0,0) -- (3,-0.6);
\draw[Lline] (0,0) -- (4,-2);
\node at (0.3,1) {$l_1$};
\node at (1.8,1.95) {$l_2$};
\node at (4,1.4) {$l_3$};
\node at (4,-0.2) {$l_4$};
\node at (3.7,-1.2) {$l_5$};
\node at (2.3,-2) {$l_6$};
\node at (0.35,-0.9) {$l_7$};
\node at (1.8,0.95) {$L_1$};
\node at (2.7,0.1) {$L_2$};
\node at (2,-0.6) {$L_3$};
\node at (2,-1.2) {$L_4$};
\end{scope}
\end{tikzpicture}
\end{center}
\caption{A multilateral illustration for the case $N = 7$.}
\label{fig:multilateraldelta}
\end{figure}
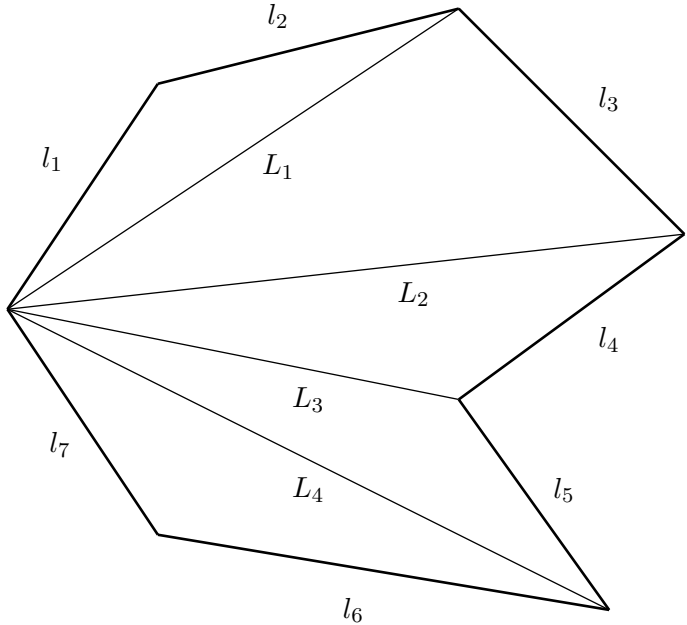

For $N=3$ the coefficients defined in \eqref{eq:Wdef} simply reduce to the Wigner $3-j$ symbol
\beq
W^{l_1l_2l_3}_{m_1m_2m_3} \equiv \( \begin{array}{ccc} l_1 & l_2 & l_3 \\ m_1 & m_2 & m_3 \end{array} \) \, .
\eeq
This quantity contains a $\{ l_1 \,\,\, l_2 \,\,\, l_3 \}$ factor, for which an alternative definition from \eqref{eq:Tdelta1} is 
\beq
\{ l_1 \,\,\, l_2 \,\,\, l_3 \} := \left\{ \begin{array}{cc} 1 & \text{if there exists a triangle with lengths} \,\,\, l_1, l_2, l_3 \\ 0 & \text{otherwise} \end{array} \right. \, ,
\eeq
hence the common name ``triangle delta". Consequently, the general coefficient \eqref{eq:Wdef} contains a $\{ l_1 \dots l_N| L_1 \dots L_{N-3} \}$ factor, defined in \eqref{eq:Mdelta} as a product of certain $\{ j_i, j_k, j_m \}$ factors. Generalizing the picture laid out in \cite{HU01a, ABPE10} for the $N=4,5$ cases,\footnote{See also section 2.4.1 of \cite{LACAS14} for the same picture, and for arbitrary $N$, but in the case of the flat sky limit where the sphere $S_2$ is replaced by $\mathbb{R}^2$.} this quantity \eqref{eq:Mdelta} can be interpreted as a ``multilateral delta" in the following way. Being a product of $N-2$ triangle deltas, it is non-zero only if all of the involved integers $l_1, \dots, l_k$ and $L_1, \dots, L_{N-3}$ form their respective triangles. Moreover, with respect to the triangle ordering in \eqref{eq:Mdelta}, one of the edges of two neighboring triangles must have the same length $L_k$, so we can picture the triangles as being connected by common edges. The resulting shape is therefore a multilateral with $N$ edges of lengths $l_1, \dots, l_N$, while the $L_1, \dots, L_{N-3}$ integers correspond to the lengths of the $N-3$ diagonals connected to the vertex where the edges $l_1$ and $l_N$ join (see figure \ref{fig:multilateraldelta}). Thus, the multilateral delta is not non-zero for {\it any} set of $l_1, \dots, l_N$ that can form a multilateral, but only for those whose diagonals also have integer length. This suggests naming the symbols $W_{\dots}^{\dots}$ defined in \eqref{eq:Wdef} the ``multilateral Wigner symbols", with the $3-j$ symbols therefore corresponding to the ``triangular" case. 

From this multilateral picture one can also directly infer that the decomposition \eqref{eq:GobofGoblN} is not unique for $N > 3$, as was already recognized in \cite{HU01a} for the case $N = 4$. Indeed, there is an ambiguity in choosing the vertex with respect to which the diagonals are drawn. This ambiguity, modulo symmetries of the $3-j$ symbols, corresponds to alternative orderings of the $(l_k, m_k)$ pairs and therefore leads us to consider permutations of the multilateral Wigner symbol indices. As shown in detail \cite{HU01a} for the $N = 4$ case, these are achieved by taking linear combinations with Wigner $6-j$ symbols, thus allowing one to relate all possible orderings.

\section{Observer terms and ensemble averaging} \label{sec:consistency}

\subsection{The issue} \label{sec:obsissue}

An important limitation in cosmology is that, observationally, we have access only to one single realization of the statistical process under consideration, the universe, and only from one single vantage point. One therefore requires some assumptions in order to relate the theoretical predictions, which are of statistical nature, to observations. In the standard approach, we consider stochastic small initial fluctuations in a homogeneous and isotropic universe, which evolve under gravity. We also assume that  their probability distribution functional is homogeneous and isotropic and that averaging over positions and directions is equivalent to ensemble average, i.e. an  ergodic hypothesis.  With these assumptions one can effectively treat different parts of the universe as different realizations and  relate the theoretical ensemble averages to observational light-cone averages. 

The requirement for the ergodic hypothesis to be applicable is the observational availability of a large enough number of source points. Indeed, the common motto in the literature is that the theorist is allowed to ensemble average a given product of fields evaluated at source positions, because the observer probes several of these positions. This viewpoint is perfectly sufficient as long as one is only interested in the lowest-order approximation where a cosmological observable is entirely determined by the value of fields at the source position. In the era of precision cosmology, however, the above approximation is no longer valid, because observations are able to capture several sub-leading effects which depend on fields evaluated on the whole line of sight from the source to the observer (e.g. weak lensing). As one approaches the observer along its past light-cone, the number of observable points decreases and therefore, one would naively infer, so does the degree of applicability of the ergodic hypothesis. The extreme case is the observer point itself, for which we can only have a single measurement. Are we then allowed to perform ensemble averages of fields at, and in the vicinity of, the observer point, as we implicitly do in \eqref{eq:Gthdef}? The aim of this section is to answer this question and its conclusion is in the affirmative. 

Before demonstrating this statement, however, let us explain in more detail why this question is relevant for precision cosmology. To that end, note first that the typical expression for a cosmological observable $\Ord$ to linear order in perturbation theory is of the form
\beq \label{eq:split}
\Ord^{(1)}(z,n) = X_{\rm o} + X_{\rm los}(z,n) + X_{\rm s}(z,n) \, , 
\eeq
where $X_{\rm o}$ and $X_{\rm s}$ denote field fluctuations evaluated at the observer and source positions, respectively, while $X_{\rm los}$ denotes an integral over fields evaluated along the background line of sight. The quantity that is relevant for comparing with observations is then, e.g., the corresponding 2-point correlation function, whose second-order contribution is
\beq \label{eq:corfunc}
G_{(2)}^{\rm th}(z,z',n,n') = \bra \Ord^{(1)}(z,n)\, \Ord^{(1)}(z',n') \ket  \, .
\eeq
Now note that the split \eqref{eq:split} is not unique, as one can always transfer quantities between the three kinds of contributions by performing integrations by parts. Moreover, since we parametrize $\Ord$ with the observed redshift $z$ and position in the sky $n$, the function $\Ord^{(1)}(z,n)$ is gauge-invariant, because the full observable $\Ord(z,n)$ has no dependence on coordinates whatsoever \cite{Yoo:2017svj}. The split \eqref{eq:split} can then be chosen such that each of the three terms are separately gauge-invariant. In practice, however, the expression \eqref{eq:split} often presents itself with a gauge-dependent observer term $X_{\rm o}$, which can therefore be set to zero with an appropriate choice of gauge, e.g. the synchronous gauge. Nevertheless, the expressions found in the literature are often in longitudinal gauge, in which case $X_{\rm o} \neq 0$. 

As already mentioned, one can consistently avoid the presence of observer terms (gauge-invariant or not) by an appropriate integration by parts, but one then needs to take into account the resulting extra terms in $X_{\rm los}$ and $X_{\rm s}$ that this manipulation introduces. Instead, the standard approach in the literature is to work in the longitudinal gauge and simply discard $X_{\rm o}$ as soon as it appears in the computation. The issue of whether one can ensemble average field values at the observer position therefore never arises in the standard practice. As for the line-of-sight terms $X_{\rm los}$, they are ensemble averaged just like the source terms $X_{\rm s}$, which implicitly assumes that ergodicity applies to all points on the light-cone, independently of how close they are to the observer point. 

It is now well understood that discarding the observer terms $X_{\rm o}$ as described above leads to several {\it qualitative} problems: 

\begin{itemize}

\item
In most cases the resulting $\Ord^{(1)}$ is not gauge-invariant under gauge transformations at the observer because one discards a gauge-dependent term $X_{\rm o}$. However, the gauge transformation of the observable only produces observer terms, by construction, which is therefore ``ok" in this approach, since one is precisely neglecting such terms to begin with. 

\item
Following the construction of section \ref{sec:ensavN}, Fourier decomposing the field configurations $\ph(t_0)$ and using statistical homogeneity to eliminate one of the $\ed^3 k$ integrals, one obtains 
\beq \label{eq:Gthofkspec}
G^{\rm th}\( t_o; \{ z_k, n_k \}_{k=1}^N \) \equiv \int \( \prod_{q=1}^{N-1} \ed^3 k_q \) \ti{G}^{\rm th}\( t_o, \{ \vec{k}_q \}_{q=1}^{N-1}; \{ z_k, n_k \}_{k=1}^N \) \, ,
\eeq
where the integrand is a combination of field $k$-spectra at $t_0$, multiplied by growth functions and $e^{i \vec{k} \cdot \vec{x}}$ factors evaluated on the light-cone. This integrand is generically divergent in the infrared, a fact which has led to confusion in the literature~\cite{Barausse:2005nf}. Moreover, the dependence on the infrared cut-off that has to be introduced to control the divergence in \eqref{eq:Gthofkspec} breaks the equivalence principle, as the observable correlation function ends up depending on the uniform gravity mode. Taking into account $X_{\rm o}$ precisely cancels the divergent terms, thus showing that they cannot be associated with physical effects and that the equivalence principle holds as one would expect \cite{Biern:2016kys,Scaccabarozzi:2018vux,Grimm:2020ays}.\footnote{For a detailed demonstration and discussion of this cancellation in the 2-point case, we refer the reader to \cite{Biern:2016kys} for the luminosity distance and to \cite{Grimm:2020ays} for the galaxy number density observables.} As discussed at the beginning of section \ref{sec:preli}, the integrand in \eqref{eq:Gthofkspec} is not itself an observable quantity, but it is interesting to see that the consideration of observer terms preserves the physical intuition in $k$-space.

\item
The sky parametrization $n$ does not generically correspond to the one of an actual observer.\footnote{See \cite{Mitsou:2019nhj} for a detailed explanation. In particular, note that while the introduction of the Sachs basis allows one to locally decompose tensors on the sky in the basis the observer uses, it does not provide the observer parametrization of the sky itself that is required in order to compute correlation functions and spectra as the observer does.} Indeed, the terms $X_{\rm o}$ contain corrections that implement the diffeomorphism relating the observer parametrization $n$ to the one induced by the local coordinate system around the observer and therefore to the gauge that is chosen. In longitudinal gauge the coordinate-induced frame in the observer's tangent space is not orthonormal in the presence of vector and tensor modes, as the one of an observer should be, so neglecting $X_{\rm o}$ leads to spurious contributions to the ``lensing" of  observables in the sky.

\end{itemize}

Nevertheless, it turns out that there is no significant {\it quantitative} problem today in neglecting $X_{\rm o}$ {\it to linear order} in perturbation theory. This is because the actual observer approximates the ensemble average in \eqref{eq:corfunc} with the SO(3) averages of the data in the sky, denoted by $\bra \dots \ket_{\rm SO(3)}$ in what follows. $X_{\rm o}$ is special with respect to this operation, because it has no angular dependence (if its spin vanishes) and can therefore be factored out. Consequently, the contribution $\bra X_{\rm o}^2 \ket_{\rm SO(3)} \equiv X_{\rm o}^2$ is a monopole term (for spin $s=0$) in the corresponding power spectrum, while the cross-terms are simply zero, e.g.
\beq
\bra X_{\rm o} X_{\rm s}(n,z) \ket_{\rm SO(3)} \equiv X_{\rm o}  \bra X_{\rm s}(n,z) \ket_{\rm SO(3)} \approx 0 \, .
\eeq
If $\Ord$ is an observable of non-zero spin $s$ (e.g. for the observer velocity, where $s=1$, or the CMB polarization, where $s = 2$), then the $\bra X_{\rm o}^2 \ket_{\rm SO(3)}$ will contribute to the first non-trivial multipole which is the $s$ multipole. But the overwhelming part of the relevant information actually lies in the rest of the spectrum, so ignoring the observer terms of cosmological observables is indeed irrelevant quantitatively.\footnote{Note that for some observables, such as galaxy number density and luminosity distance, the standard analysis is performed with correlation functions instead of spectra, in which case all multipoles contribute to the result and the lower-order ones are not subtracted. In that case, it has been shown that observer terms can be of the same order as other relativistic corrections \cite{Biern:2017bzo,Scaccabarozzi:2018vux}.} 

The caveat of the previous argumentation is that this is only true {\it to linear order in perturbation theory}. Indeed, to second order the solution of the observable will have cross-terms of the schematic form\footnote{For a concrete example, see equations (4.2) to (4.5) of \cite{Fanizza:2018qux} for the second-order part of the redshift perturbation.}
\beq
\Ord^{(2)}(z,n) \supset X_{\rm o} X_{\rm s}(z,n) + X_{\rm o} X_{\rm los}(z,n) + \dots 
\eeq
implying terms of the form 
\beq
G^{\rm th}_{(3)}(z,z',n,n') \supset \bra X_{\rm o} X_{\rm s}(z,n)\, X_{\rm s}(z',n') \ket + \dots
\eeq
for the next order of the 2-point correlation function. Importantly, these contributions will affect {\it all} multipoles of the third-order power spectrum with an amplitude that is a priori comparable to the standard terms such as $\sim \bra X^2_{\rm s} X_{\rm los} \ket$, etc. Moreover, similar terms will also appear in higher-order correlation functions. Thus, in the era of precision cosmology one can no longer ignore the observer terms $X_{\rm o}$, in the usual gauge choices that are considered, meaning that it is relevant to discuss their inclusion inside ensemble averages.

\subsection{Relating theory and observation}  \label{sec:relthob}

We now wish to derive the relation between the observational and theoretical $N$-point functions of observables \eqref{eq:Gobdef} and \eqref{eq:Gthdef}. We already have equation \eqref{eq:ensofGob}, but this is a theoretical relation, as it involves ensemble averaging, and simply states that $G^{\rm ob}$ is a full unbiased estimator of $G^{\rm th}$. Rather, we are interested in relating the two $N$-point functions as one does in practice, i.e. we need $G^{\rm ob}$ to be evaluated on a definite field configuration, and this is achieved using the ergodic hypothesis. In particular, we want to see how the latter enters the derivation and how it affects the issue of the single observer point. 

We start by recalling subsection \ref{sec:ensavN} and consider the set of 3-dimensional field configurations $\ph_{\frak{a}}(t_0)$ over which we sum when performing the ensemble average in \eqref{eq:Ptmom}. Given the invariance of $P[\ph(t_0)]$ under the isometry group, it is convenient to partition this ensemble into equivalence classes, where two field configurations are deemed equivalent if they can be related by an isometry. For the sake of simplicity, let us consider here the case of flat background space, so that the isometries form the Euclidean group $\mathbb{E} := {\rm SO(3)} \ltimes \Rs^3$, and let us also use the trivial Cartesian coordinates. Any element of a given equivalence class can be described as an isometry of some fixed representative $\hat{\ph}_{\frak{a}}(t_0)$ 
\beq
\ph_{\frak{a},R,\vec{c}}(t_0,\vec{x}) = M_{\frak{a}}^{\,\,\frak{b}}(R)\, \hat{\ph}_{\frak{b}}(t_0,R \vec{x} + \vec{c}) \, ,
\eeq
where $R$ is a rotation matrix, $\vec{c}$ a translation vector and $M_{\frak{a}}^{\,\,\frak{b}}$ is the matrix that rotates  tensor indices in $\frak{a}$. We can therefore split the functional integration in \eqref{eq:Ptmom} into an integral over the elements of a given class followed by an integral over all possible classes. By the latter we mean an integral over suitably chosen representatives $\hat{\ph}_{\frak{a}}(t_0)$ such that the corresponding functional integral is well-defined.\footnote{The existence of such a splitting of the integration is a non-trivial mathematical assertion, whose proof, if possible, would go beyond the scope of this paper.} Since the pdf is constant over all representatives of a given class, the integral in \eqref{eq:Ptmom} becomes
\beq
\bra \ph_{\frak{a}_1}(t_0,\vec{x}_1) \dots \ph_{\frak{a}_n}(t_0,\vec{x}_n) \ket \equiv \frac{\int D \hat{\ph}(t_0) \, P[\hat{\ph}(t_0)] \, \bra \hat{\ph}_{\frak{a}_1}(t_0,\vec{x}_1) \dots \hat{\ph}_{\frak{a}_n}(t_0,\vec{x}_n) \ket_{\mathbb{E}}}{\int D \hat{\ph}(t_0) \, P[\hat{\ph}(t_0)]} \, , \label{eq:part}
\eeq
where we now integrate only over the set of representatives, and
\beq
\bra X[\ph(t_0)] \ket_{\mathbb{E}} := \int \frac{\ed^3 c}{V} \int \frac{\ed R}{8\pi^2} \, X[M_{\frak{a}}^{\,\,\frak{b}}(R)\, \ph_{\frak{b}}(t_0,R \vec{x} + \vec{c})]  \, .
\eeq
is the average over the Euclidean group action over the field configurations, $V$ is the total volume and $\ed R$ is the SO(3) Haar measure. As one may expect, the ensemble average therefore contains a purely geometric average over the symmetry group of the pdf. 

Apart from subsets of measure zero, the configurations $\hat{\ph}(t_0)$ appearing in the integral \eqref{eq:part} have a rich spatial dependence. In particular, they are non-periodic functions which therefore probe a large variety of local field profiles for large enough $V$. In the $V \to \infty$ limit, which we consider here, the ergodic hypothesis states that this probing is thorough enough to make the Euclidean average in \eqref{eq:part} independent of the configuration $\hat{\ph}(t_0)$. As a result, that average factorizes out of the integral, thus yielding the relation
\beq \label{eq:erg1}
\bra \ph_{\frak{a}_1}(t_0,\vec{x}_1) \dots \ph_{\frak{a}_n}(t_0,\vec{x}_n) \ket \os{\rm erg.}{=} \bra \ph_{\frak{a}_1}(t_0,\vec{x}_1) \dots \ph_{\frak{a}_n}(t_0,\vec{x}_n) \ket_{\mathbb{E}} \, .
\eeq
Now on the right-hand side it is a definite, although generic field configuration that is considered. We can then generalize this manipulation straightforwardly to the case of the FCFs \eqref{eq:fieldG}, since the Euclidean group action is independent of the time variable $t$, and therefore also to the case of the TCF \eqref{eq:Gthdef}
\bea
G^{\rm th} \( t_o; \{ z_k, n_k \}_{k=1}^N \) & \os{\rm erg.}{=} & \Bra \prod_{k=1}^N \Ord\( x_o; z_k, n_k  \)[\Phi] \Ket_{\mathbb{E}} \label{eq:Gthmanip1} \\
 & \equiv & \int \frac{\ed^3 c}{V} \int \frac{\ed R}{8\pi^2} \, \prod_{k=1}^N \Ord\( x_o; z_k, n_k  \) \[M_{\frak{a}}^{\,\,\frak{b}}(R)\, \ph_{\frak{b}}(t,R \vec{x} + \vec{c}) \]   \, . \nn
\eea
where now we act with $\mathbb{E}$ directly on the 4-dimensional fields $\ph_{\frak{a}}(x)$ instead of the $\ph_{\frak{a}}(t_0,\vec{x})$. Note also that we have not specified the $\bar{\Phi}$ dependence in \eqref{eq:Gthmanip1} for simplicity. The above expression allows us to make contact with the OCFs \eqref{eq:Gobdef}. Let us first redefine the dummy variable $\vec{c} \to \vec{c} - R \vec{x}_o$ in \eqref{eq:Gthmanip1} and let us also define the notation for shifted fields
\beq
\ph_{\frak{a},\vec{c}}(t,\vec{x}) := \ph_{\frak{a}}(t,\vec{x} + \vec{c}) \, ,
\eeq
to get 
\beq  \label{eq:Gthmanip2}
G^{\rm th} \( t_o; \{ z_k, n_k \}_{k=1}^N \) \os{\rm erg.}{=} \int \frac{\ed^3 c}{V} \int \frac{\ed R}{8\pi^2} \, \prod_{k=1}^N \Ord\( x_o; z_k, n_k  \) \[ M_{\frak{a}}^{\,\,\frak{b}}(R)\, \ph_{\frak{b},\vec{c}}(t,R (\vec{x} - \vec{x}_o)) \] \, .
\eeq
We next observe that, for a given value of $\vec{c}$, the fields $\ph_{\frak{a},\vec{c}}$ are rotated by $R^{-1}$ around $\vec{x}_o$ in \eqref{eq:Gthmanip2}. But rotating the fields around $\vec{x}_o$ is tantamount to rotating the observed angles $n$ in the opposite direction, so the SO(3) average on the fields translates into an OCF
\bea
G^{\rm th} \( t_o; \{ z_k, n_k \}_{k=1}^N \) & \os{\rm erg.}{=} & \int \frac{\ed^3 c}{V} \int \frac{\ed R}{8\pi^2} \, \prod_{k=1}^N \Ord\( x_o; z_k, R^{-1} n_k  \)[\ph_{\frak{a},\vec{c}}(t, \vec{x} - \vec{x}_o)] \nn \\
 & \equiv & \int \frac{\ed^3 c}{V}\, G^{\rm ob}\( x_o; \{ z_k, n_k \}_{k=1}^N \)[\ph_{\frak{a},\vec{c}}(t, \vec{x} - \vec{x}_o)] \nn \\
 & \equiv & \int \frac{\ed^3 c}{V}\, G^{\rm ob}\( x_o; \{ z_k, n_k \}_{k=1}^N \)[\ph_{\frak{a}}(t, \vec{x} - \vec{x}_o + \vec{c})]  \, . \label{eq:Gthmanip3}
\eea
Similarly, translating the fields by $\vec{x}_o - \vec{c}$ can be equivalently expressed as shifting the observer by $\vec{c} - \vec{x}_o$, so we finally obtain (after renaming $\vec{c} \to \vec{x}_o$)
\beq \label{eq:erg}
G^{\rm th} \( t_o; \{ z_k, n_k \}_{k=1}^N \) \os{\rm erg.}{=} \frac{1}{V} \int \ed^3 x_o \, G^{\rm ob} \( t_o, \vec{x}_o; \{ z_k, n_k \}_{k=1}^N \)  \, .
\eeq
i.e. the average over the action of the Euclidean group on the fields $\ph_{\frak{a}}$ is equivalent to averaging over all observer reference frames, as in $G^{\rm ob}$, but {\it also} over all observer positions. This is simply because we have imposed both statistical isotropy {\it and} homogeneity. For instance, the CMB maps one would obtain from another vantage point of the universe $\vec{x}'_o$ would be different from the ones we observe on earth today, while the theoretical power spectrum which we calculate is the same for all vantage points. Since the relation between the correlation function and their associated spectra is linear, we have that \eqref{eq:erg} simply translates into 
\beq  \label{eq:ergh}
G^{\rm th}_{l_1 \dots l_N| L_1 \dots L_{N-3}} \( t_o; \{ z_k \}_{k=1}^N \) \os{\rm erg.}{=} \frac{1}{V} \int \ed^3 x_o \, G^{\rm ob}_{l_1 \dots l_N| L_1 \dots L_{N-3}} \( t_o, \vec{x}_o; \{ z_k \}_{k=1}^N \)  \, .
\eeq
Equations \eqref{eq:erg} and \eqref{eq:ergh} are the central result of this section. The important thing to understand from them is that the application of the ergodic hypothesis singles out a definite field configuration on the right-hand side, but it does not single out the position of the actual observer, since all possible such positions are taken into account equally inside an average. Therefore, the applicability of the ergodic hypothesis depends on the representative field configuration which we choose on the right-hand side of \eqref{eq:erg}, {\it completely independently of how this configuration is probed observationally}. In particular, as stated in the step in which ergodicity is actually used \eqref{eq:erg1}, what we need is that field configuration to have a rich profile, which is a decent assumption to make about the configuration of the actual universe.  

What we also see, however, is that the use of ergodicity is necessary, but not sufficient to relate theory and observations yet, because the information from several $\vec{x}_o$ that is required in \eqref{eq:erg} and \eqref{eq:ergh} is not observationally available. Rather, the step which actually singles out the observer point is the approximation of the $\vec{x}_o$ averages in the right hand-sides of \eqref{eq:erg} and \eqref{eq:ergh} by the value at the actual observer position
\beq \label{eq:stderg}
G^{\rm th}\( t_o; \{ z_k, n_k \}_{k=1}^N \) \os{\rm erg.}{\approx} G^{\rm ob}\( x_o;\{ z_k, n_k \}_{k=1}^N \)  \, ,  
\eeq
and
\beq \label{eq:stdergh}
G^{\rm th}_{l_1 \dots l_N| L_1 \dots L_{N-3}} \( t_o; \{ z_k \}_{k=1}^N \) \os{\rm erg.}{\approx} G^{\rm ob}_{l_1 \dots l_N| L_1 \dots L_{N-3}} \( x_o; \{ z_k \}_{k=1}^N \) \, ,
\eeq
respectively. Approximating an average over some set by a single value within that set is of course the crudest possible estimator of that average. One expects this estimation to be accurate enough only if the values of the considered set exhibit a relatively small dispersion around the average value. But this dispersion in $\vec{x}_o$ space cannot be measured for the very same reason that brought us to this approximation in the first place, i.e. the observational unavailability of data at other observation points. Nevertheless, as we will see in the next subsection, the variance of $G^{\rm ob}$ with respect to the $\vec{x}_o$ dependence is equal to its theoretical analogue, the statistical or ``cosmic'' variance, through the ergodic hypothesis. Thus, the error one makes in \eqref{eq:stderg} or \eqref{eq:stdergh} is precisely cosmic variance, which is indeed negligible on small enough scales (see section \ref{sec:clth}).

Equations \eqref{eq:stderg} and \eqref{eq:stdergh} allow to relate theory and observation, as both sides can be computed in their respective domains. Finally, another way of interpreting the result \eqref{eq:erg} is by combining it with \eqref{eq:ensofGob} to obtain
\beq  \label{eq:enseqtrav}
\Bra G^{\rm ob}\( x_o; \{ z_k, n_k \}_{k=1}^N \) \Ket \os{\rm erg.}{=} \frac{1}{V} \int \ed^3 x_o \,G^{\rm ob}\( x_o;\{ z_k, n_k \}_{k=1}^N \)  \, , 
\eeq
i.e. the ergodic hypothesis and statistical homogeneity equate the ensemble averaging with the translational averaging of a generic configuration. Thus, under these assumptions, while estimating $G^{\rm th}$ with $G^{\rm ob}$ first appears as a double approximation, i.e. the measurement of a single universe realization {\it and} from a single viewpoint, these two turn out to be the very same approximation.

\subsection{Unambiguous covariance matrix and cosmic variance} \label{sec:unamb}

Equation \eqref{eq:enseqtrav} shows that, assuming ergodicity and statistical homogeneity, ensemble averaging $G^{\rm ob}$ is tantamount to averaging it over $\vec{x}_o$ with a generic configuration of the fields. The last open question is therefore whether these two types of averaging are also equal when it comes to the variance associated with the estimation \eqref{eq:stderg}. The statistical covariance matrix is 
\bea
{\rm Cov}_{\rm stat.}(t_o; \al_N; \al'_{N'}) & := & \Bra \[ G^{\rm ob}(x_o; \al_N) - G^{\rm th}(\al_N) \] \[ G^{\rm ob}(x_o; \al'_{N'}) - G^{\rm th}(\al'_{N'}) \] \Ket \nn \\
 & \equiv & \bra G^{\rm ob}(x_o; \al_N)\, G^{\rm ob}(x_o; \al'_{N'}) \ket - G^{\rm th}(\al_N)\, G^{\rm th}(\al'_{N'})  \, , \label{eq:sistat}
\eea
where $\al_N$ collectively denotes the set $\{ z_k, n_k \}_{k=1}^N$ for notational simplicity, while the spatial one is
\bea
{\rm Cov}_{\rm spat.}(t_o; \al_N; \al'_{N'}) & := & \frac{1}{V} \int \ed^3 x_o \[ G^{\rm ob}(x_o; \al_N) - G^{\rm th}(\al_N) \] \[ G^{\rm ob}(x_o; \al'_{N'}) - G^{\rm th}(\al'_{N'}) \] \nn \\
 & \os{\rm erg.}{=} & \frac{1}{V} \int \ed^3 x_o \, G^{\rm ob}(x_o; \al_N) \, G^{\rm ob}(x_o; \al'_{N'}) - G^{\rm th}(\al_N)\, G^{\rm th}(\al'_{N'})  \, . \label{eq:sigeo}
\eea
The former leads to the notion of cosmic variance, so, if both of them are the same, we would have shown that our indiscriminate use of ensemble averaging leads to no new uncertainties between theory and observation. This is simply achieved by noting that a product of two OCFs can be expressed as a partial SO(3) average of a single OCF
\bea
G^{\rm ob}(x_o; \al_N) \, G^{\rm ob}(x_o; \al'_{N'}) & \equiv & \frac{1}{(8\pi^2)^2} \int \ed R \, \int \ed R'\, \prod_{k=1}^N \Ord(x_o; z_k, R^{-1} n_k) \label{eq:prodGob} \\
 & & \hspace{3.6cm} \times \prod_{k'=1}^{N'} \Ord(x_o; z'_{k'}, R'^{-1}n'_{k'}) \nn \\
 & \os{R \to R R'}{\equiv} & \frac{1}{(8\pi^2)^2} \int \ed R \, \int \ed R'\, \prod_{k=1}^N \Ord(x_o; z_k, R'^{-1} R^{-1} n_k) \nn \\
 & & \hspace{3.6cm} \times \prod_{k'=1}^{N'} \Ord(x_o; z'_{k'}, R'^{-1}n'_{k'}) \nn \\
 & \equiv & \frac{1}{8\pi^2} \int \ed R \, G^{\rm ob}\( x_o; \{ z_k, R^{-1} n_k \}_{k=1}^N, \{ z'_{k'}, n'_{k'} \}_{l=1}^{N'} \) \, .    \nn
\eea 
Inserting this in \eqref{eq:sistat} and \eqref{eq:sigeo} and using \eqref{eq:erg} and \eqref{eq:ensofGob} we find that both covariances are equal
\beq \label{eq:sieq}
{\rm Cov}_{\rm spat.} \os{\rm erg.}{=} {\rm Cov}_{\rm stat.} =: {\rm Cov} \, ,
\eeq
and that
\beq \label{eq:si}
{\rm Cov} \( t_o; \al_N; \al'_M \) \equiv \frac{1}{8\pi^2} \int \ed R \, G^{\rm th}\( \{ z_k, R^{-1} n_k \}_{k=1}^N, \{ z'_l, n'_l \}_{l=1}^M \) - G^{\rm th}(\al_N)\, G^{\rm th}(\al'_M) \, .
\eeq
Thus, the absolute ``1-sigma" error of the estimator \eqref{eq:stderg}
\beq \label{eq:sidef}
\Si(t_o; \al_N) := \sqrt{{\rm Cov}(t_o; \al_N; \al_N)} \, ,
\eeq
is unambiguously cosmic variance. We see that our derivation provides a refined understanding of that notion, which is not usually expressed in cosmology textbooks. Given our statistical assumptions, cosmic variance is the error due to the fact that we observe a single realization of the universe {\it and} from a single vantage point $\vec{x}_o$. In accordance with the discussion of the previous subsection, if either of these two conditions were dropped, then there would be no cosmic variance. On one hand, if we had simultaneous access to the data of a single realization from all possible $\vec{x}_o$, then the ergodic hypothesis \eqref{eq:erg} would allow us to match the theoretical predictions exactly.\footnote{Here we neglect the fact that this information would also require a time $\sim V^{1/3}$ to be collected by a main observer and therefore analyzed.} On the other hand, if we could observe all possible universe realizations, even from a single viewpoint $\vec{x}_o$, then we would be able to compute directly the theoretical $N$-point functions, which are independent of $\vec{x}_o$. In the first case we are technically setting $\Si_{\rm spat.} \to 0$, whereas in the second one it is rather $\Si_{\rm stat.} \to 0$. 

In the next section we will see that, although cosmic variance is always non-zero in practice, the associated relative uncertainty on the spectra must tend to zero when $l_k \to \infty$ as a consequence of statistical isotropy and the central limit theorem. Technically, this corresponds to $\Si_{\rm stat.}$ becoming negligible compared to $G^{\rm th}$ for large enough $l_k$ values, with statistical homogeneity and the ergodic hypothesis then implying that so does $\Si_{\rm spat.}$, as we just saw in \eqref{eq:sieq}. This is the reason why, in practice, one can still obtain statistical cosmological information even from a single observational point.

\section{Covariance in $l$-space} \label{sec:extra}

\subsection{The covariance matrix of angular $N$-point spectra}

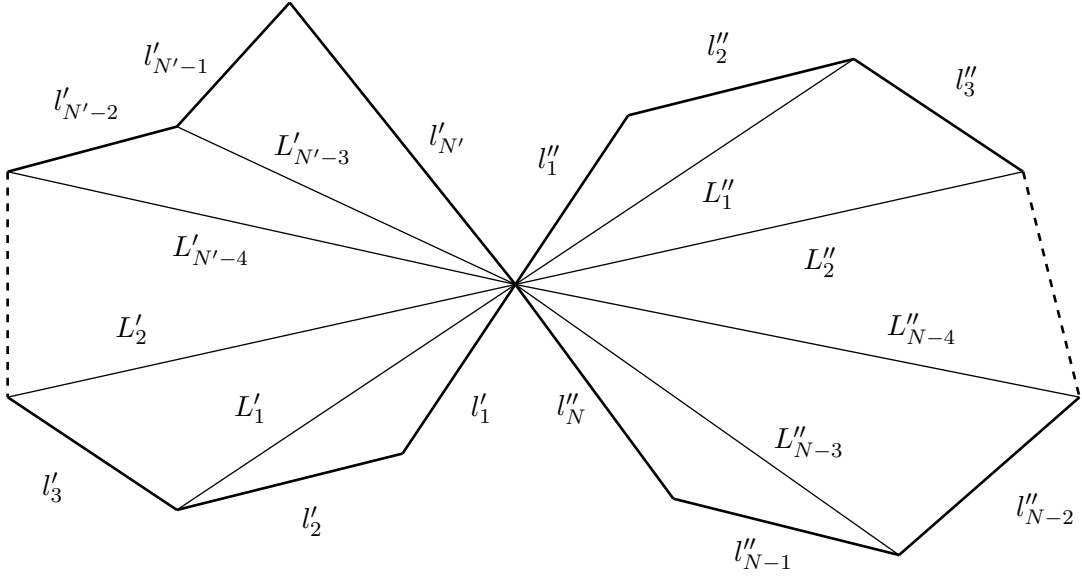
\begin{figure}[h!] 
\hspace*{0.6cm} \begin{tikzpicture}[scale = 1.5]
\begin{scope}
\draw[lline] (0,0) -- (1,1.5);
\draw[lline] (1,1.5) -- (3,2);
\draw[lline] (3,2) -- (4.5,1);
\draw[dline] (4.5,1) -- (5,-1);
\draw[lline] (5,-1) -- (3.4,-2.4);
\draw[lline] (3.4,-2.4) -- (1.4,-1.9);
\draw[lline] (1.4,-1.9) -- (0,0);
\draw[Lline] (0,0) -- (3,2);
\draw[Lline] (0,0) -- (4.5,1);
\draw[Lline] (0,0) -- (5,-1);
\draw[Lline] (0,0) -- (3.4,-2.4);
\node at (0.3,1.1) {$l''_1$};
\node at (1.8,2.1) {$l''_2$};
\node at (4,1.8) {$l''_3$};
\node at (0.5,-1.1) {$l''_{N}$};
\node at (2.2,-2.4) {$l''_{N-1}$};
\node at (4.7,-2) {$l''_{N-2}$};
\node at (1.8,0.8) {$L''_1$};
\node at (2.7,0.2) {$L''_2$};
\node at (2.6,-1.4) {$L''_{N-3}$};
\node at (3.6,-0.4) {$L''_{N-4}$};
\draw[lline] (0,0) -- (-2,2.5);
\draw[lline] (-2,2.5) -- (-3,1.4);
\draw[lline] (-3,1.4) -- (-4.5,1);
\draw[dline] (-4.5,1) -- (-4.5,-1);
\draw[lline] (-4.5,-1) -- (-3,-2);
\draw[lline] (-3,-2) -- (-1,-1.5);
\draw[lline] (-1,-1.5) -- (0,0);
\draw[Lline] (0,0) -- (-3,1.4);
\draw[Lline] (0,0) -- (-4.5,1);
\draw[Lline] (0,0) -- (-4.5,-1);
\draw[Lline] (0,0) -- (-3,-2);
\node at (-0.6,1.3) {$l'_{N'}$};
\node at (-3,2) {$l'_{N'-1}$};
\node at (-3.8,1.6) {$l'_{N'-2}$};
\node at (-0.3,-1.1) {$l'_1$};
\node at (-1.8,-2.1) {$l'_2$};
\node at (-4.1,-1.8) {$l'_3$};
\node at (-1.8,1.2) {$L'_{N'-3}$};
\node at (-2.7,0.3) {$L'_{N'-4}$};
\node at (-2.35,-1.1) {$L'_1$};
\node at (-3.4,-0.4) {$L'_2$};
\end{scope}
\end{tikzpicture}
\caption{The type of diagram composing the covariance matrix of the $N$-point and $N'$-point spectra.}
\label{fig:covariancediagram}
\end{figure}

From now on we drop the dependence on $t_o$ for notational simplicity. Here we wish to derive the expression for the covariance matrix of the reduced $N$-point and $N'$-point spectra, i.e. the harmonic analogue of \eqref{eq:si}. To that end, we first note that ${\rm Cov} \( t_o; \al_N; \al'_{N'} \)$ is a function on $S_2^N \times S_2^{N'}$ that is invariant under independent rotations on each of these two components, so we can decompose each of the corresponding arguments in the basis \eqref{eq:GenYbasis}. The harmonic components are then found by projecting twice
\bea
 & & {\rm Cov}_{l_1 \dots l_N| L_1 \dots L_{N-3}; l'_1 \dots l'_{N'}| L'_1 \dots L'_{N'-3}}\( \{ z_k \}_{k=1}^N ; \{ z'_{k'} \}_{k'=1}^{N'} \) \nn \\
 & := & \int \(  \prod_{k=1}^N \frac{\ed \Om_k}{4\pi} \) \(  \prod_{k'=1}^{N'} \frac{\ed \Om'_{k'}}{4\pi} \) Y^*_{l_1 \dots l_N| L_1 \dots L_{N-3}}\( \{ n_k \}_{k=1}^N \) Y^*_{l'_1 \dots l'_{N'}| L'_1 \dots L'_{N'-3}}\( \{ n'_{k'} \}_{k'=1}^{N'} \) \nn \\
 & & \hspace{1cm} \times {\rm Cov}\( \{ z_k, n_k \}_{k=1}^N ; \{ z'_{k'}, n'_{k'} \}_{k'=1}^{N'} \) \, . 
\eea
To express this in terms of the reduced theoretical spectra, we use \eqref{eq:si} and decompose the TCFs in the basis \eqref{eq:GenYbasis}. By proceeding exactly as in appendix \ref{sec:hdec} to eliminate the SO(3) average in \eqref{eq:si}, and using the orthonormality relations \eqref{eq:Ynorm} and \eqref{eq:Wortho}, we find
\bea
& & {\rm Cov}_{l_1 \dots l_N| L_1 \dots L_{N-3}; l'_1 \dots l'_{N'}| L'_1 \dots L'_{N'-3}}\( \{ z_k \}_{k=1}^N ; \{ z'_{k'} \}_{k'=1}^{N'} \) \nn \\
 & \equiv & W^{l_1 \dots l_N|L_1 \dots L_{N-3}}_{m''_1 \dots m''_N} W^{l''_1 \dots l''_N l'_1 \dots l'_{N'} | L''_1 \dots L''_{N+N'-3}}_{m''_1 \dots m''_N m'_1 \dots m'_{N'}} W^{l'_1 \dots l'_{N'}| L'_1 \dots L'_{N'-3}}_{m'_1 \dots m'_{N'}} \nn \\
 & & \times \, G^{\rm th}_{l''_1 \dots l''_N l'_1 \dots l'_{N'}| L''_1 \dots L''_{N+N'-3}}\( \{ z_k \}_{k=1}^N, \{ z'_{k'} \}_{k'=1}^{N'} \) \nn \\
 & & - \, G^{\rm th}_{l_1 \dots l_N | L_1 \dots L_{N-3}}\( \{ z_k \}_{k=1}^N \) G^{\rm th}_{l'_1 \dots l'_{N'} | L'_1 \dots L'_{N'-3}}\( \{ z'_{k'} \}_{k'=1}^{N'} \) \, .
\eea
Now the second product of Wigner symbols can be simplified using \eqref{eq:3jsum1} iteratively, then \eqref{eq:3jsum2} and finally \eqref{eq:3j0} to find
\bea
 & & \sum_{m'_k} W^{l''_1 \dots l''_N l'_1 \dots l'_{N'} | L''_1 \dots L''_{N+N'-3}}_{m''_1 \dots m''_N m'_1 \dots m'_{N'}} W^{l'_1 \dots l'_{N'}| L'_1 \dots L'_{N'-3}}_{m'_1 \dots m'_{N'}} \label{eq:WWinter} \\
 & \equiv & \{ l'_1 \dots l'_{N'} | L'_1 \dots L'_{N'-3} \} \de_{l''_N L''_{N-2}} \de_{0 L''_{N-1}} \de_{l'_1 L''_N} \( \prod_{k=1}^{N'-3} \de_{L'_k L''_{N+k}} \) W^{l''_1 \dots l''_N | L''_1 \dots L''_{N-3}}_{m''_1 \dots m''_N} \, . \nn
\eea
We therefore obtain
\bea
& & {\rm Cov}_{l_1 \dots l_N| L_1 \dots L_{N-3}; l'_1 \dots l'_{N'}| L'_1 \dots L'_{N'-3}}\( \{ z_k \}_{k=1}^N ; \{ z'_{k'} \}_{k'=1}^{N'} \) \nn \\
 & \equiv & A_{l_1 \dots l_N|L_1 \dots L_{N-3}}^{l''_1 \dots l''_N|L''_1 \dots L''_{N-3}} G^{\rm th}_{l''_1 \dots l''_N l'_1 \dots l'_{N'}| L''_1 \dots L''_{N-3} l''_N 0 l'_1 L'_1 \dots L'_{N'-3}}\( \{ z_k \}_{k=1}^N, \{ z'_{k'} \}_{k'=1}^{N'} \) \nn \\
 & & - \, G^{\rm th}_{l_1 \dots l_N | L_1 \dots L_{N-3}}\( \{ z_k \}_{k=1}^N \) G^{\rm th}_{l'_1 \dots l'_{N'} | L'_1 \dots L'_{N'-3}}\( \{ z'_{k'} \}_{k'=1}^{N'} \) \, , \label{eq:gencovl}
\eea
where we have defined the following contraction of multilateral Wigner symbols of equal order
\beq \label{eq:Adef}
A^{l_1 \dots l_N|L_1 \dots L_{N-3}}_{l'_1 \dots l'_N|L'_1 \dots L'_{N-3}} := W^{l_1 \dots l_N|L_1 \dots L_{N-3}}_{m_1 \dots m_N} W^{l'_1 \dots l'_N | L'_1 \dots L'_{N-3}}_{m_1 \dots m_N} \, ,
\eeq
which is rotationally-invariant, as it does not depend on $m_k$ indices. From \eqref{eq:gencovl} we see that the covariance matrix of the $N$-point and $N'$-point spectra is controlled by the squeezed $(N+N')$-point spectrum, as depicted in figure \ref{fig:covariancediagram}, which arises from gluing together an $N$-point and an $N'$-point diagram. Therefore, the squeezed $(N+N')$-point spectrum here is already proportional to $\{ l'_1 \dots l'_{N'} | L'_1 \dots L'_{N'-3} \}$, which is why we were able to discard that factor coming from \eqref{eq:WWinter}. 

The coefficients defined in \eqref{eq:Adef} form a symmetric square matrix in the space of reduced $N$-point spectra and in \eqref{eq:gencovl} we see that this matrix acts on the $N$-point component of the squeezed $(N+N')$-point spectrum. Comparing with the analogous expression for the correlation functions \eqref{eq:si}, we thus see that the $A$ matrix \eqref{eq:Adef} implements the operation of partial SO(3) average on invariant functions in harmonic space. Finally, note that one can extend the expression \eqref{eq:gencovl} to the cases $N,N'= 1,2$ by using a redundant notation analogous to Eqs. \eqref{eq:Gob2spec}, \eqref{eq:Yll2spec} and \eqref{eq:Wdef12}.

\subsection{Cosmic variance in the large $l_k$ limit} \label{sec:clth}

Let us consider the case $N = N'$, where the covariance matrix quantifies the typical error of the $G^{\rm th} \approx G^{\rm ob}(\vec{x}_o)$ approximation that is cosmic variance, so that the $l$-space analogue of \eqref{eq:sidef} reads
\beq
\Si_{l_1 \cdots l_N| L_1 \cdots L_{N-3}}\( \{ z_k \}_{k=1}^N \) := \sqrt{{\rm Cov}_{l_1 \dots l_N| L_1 \dots L_{N-3}; l_1 \dots l_{N}| L_1 \dots L_{N-3}}\( \{ z_k \}_{k=1}^N ; \{ z_{k} \}_{k=1}^{N} \)} \, .
\eeq
We now note that this quantity has a simple asymptotic behavior at large $l_k$, which is completely determined by the central limit theorem~\cite{Feller}. Indeed, thanks to the assumption of statistical isotropy, the $\Ord_{lm}(\vec{x}_o)$ for a given value of $l$ are $2 l + 1$ independent and equally distributed random variables. Thus, the product $\prod_{k=1}^N \Ord_{l_k m_k}(\vec{x}_o)$ for given $l_k$ values consists of $\Ord \( \prod_{k=1}^N l_k \)$ independent and equally distributed random variables. The observational spectrum component $G^{\rm ob}_{l_1 \cdots l_N| L_1 \cdots L_{N-3}}(\vec{x}_o)$ then corresponds to a weighted sum of these products over their $m_k$ indices \eqref{eq:GobNl}. However, since the multilateral Wigner symbol \eqref{eq:Wdef} is zero unless $\sum_{k=1}^N m_k = 0$, the sum actually only contains $\Ord \( \prod_{k=1}^{N-1} l_k \)$ independent and equally distributed random variables. Finally, noting that $G^{\rm ob}(\vec{x}_o)$ is technically a sample average and $G^{\rm th}$ is its ensemble average, the central limit theorem~\cite{Feller} states that
\beq \label{eq:Silgen}
\left. \Si_{l_1 \cdots l_N| L_1 \cdots L_{N-3}} \frac{}{} \right|_{l_k \gg 1} \propto \frac{1}{\sqrt{\prod_{k=1}^{N-1} l_k}} \, G^{\rm th}_{l_1 \cdots l_N| L_1 \cdots L_{N-3}}  \, .
\eeq 
Consequently, the relative cosmic variance tends to zero as $l_k \to \infty$. In particular, this means that $G^{\rm ob}_{l_1 \cdots l_N| L_1 \cdots L_{N-3}}(\vec{x}_o)$ tends to the $\vec{x}_o$-independent result $G^{\rm th}_{l_1 \cdots l_N| L_1 \cdots L_{N-3}}$, without the need to perform the average over $\vec{x}_o$ as in \eqref{eq:ergh}, thus justifying the approximation \eqref{eq:stderg} for small enough scales. This is therefore how one can obtain accurate statistical cosmological information from a single vantage point for sufficiently large $l_k$. Following the discussion at the end of section \ref{sec:consistency}, this loss of the $\vec{x}_o$ information in the $l_k \to \infty$ limit is a consequence of statistical homogeneity and the ergodic hypothesis.

\section{Connected angular $N$-point spectra}  \label{sec:connected}

\subsection{Definitions, modified ergodic relation and covariance matrices}

Here we focus on the part that is usually of most interest: the {\it connected} component of a given correlation function or associated spectrum. This is a non-linear combination of the full statistics $G$, so one should expect the linear ergodic relations \eqref{eq:erg} and \eqref{eq:ergh} to be modified. Here we will illustrate this modification by considering the case of the connected 2,3, and 4-point functions. We start by defining the observable fluctuations, in the observational and theoretical cases, as the deviation of the observable from the respective 1-point functions
\bea
\De^{\rm ob} \Ord(\vec{x}_o;z,n) & := & \Ord(\vec{x}_o;z,n) - G^{\rm ob}(\vec{x}_o;z) \, , \\
\De^{\rm th} \Ord(\vec{x}_o;z,n) & := & \Ord(\vec{x}_o;z,n) - G^{\rm th}(z) \, ,
\eea
or, in terms of the harmonic components,
\bea 
\De^{\rm ob} \Ord_{lm}(\vec{x}_o;z) & := & \Ord_{lm}(\vec{x}_o;z) - \de^0_l \de^0_m G^{\rm ob}(\vec{x}_o;z) \, , \label{eq:Defdefs} \\
\De^{\rm th} \Ord_{lm}(\vec{x}_o;z) & := & \Ord_{lm}(\vec{x}_o;z) - \de_l^0 \de^0_m G^{\rm th}(z) \, .
\eea
We note in particular that, by construction, the observational monopole is identically zero, while the theoretical one captures precisely the difference between the two 1-point functions
\bea
\De^{\rm ob} \Ord_{00}(\vec{x}_o; z) & \equiv & G^{\rm ob}(\vec{x}_o,z) - G^{\rm ob}(\vec{x}_o,z) \equiv 0 \, , \label{eq:obiden} \\
\De^{\rm th} \Ord_{00}(\vec{x}_o; z) & \equiv & G^{\rm ob}(\vec{x}_o,z) - G^{\rm th}(z) \neq 0 \, . \label{eq:thiden}
\eea 
The connected 2,3,4-point functions can then be defined by
\bea
C^{\star}(z_1,z_2, n_1, n_2) & := & \Bra \De^{\star} \Ord(z_1,n_1)\, \De^{\star} \Ord(z_2,n_2) \Ket_{\star} \, , \label{eq:Pob} \\
B^{\star}(z_1, z_2, z_3, n_1, n_2, n_3) & := & \Bra \De^{\star} \Ord(z_1,n_1)\, \De^{\star} \Ord(z_2,n_2) \, \De^{\star} \Ord(z_3,n_3) \Ket_{\star} \, , \\
T^{\star}(z_1, z_2, z_3,z_4, n_1, n_2, n_3, n_4) & := & \Bra \De^{\star} \Ord(z_1,n_1)\, \De^{\star} \Ord(z_2,n_2) \, \De^{\star} \Ord(z_3,n_3) \, \De^{\star} \Ord(z_4,n_4) \Ket_{\star} \label{eq:Tobdef}  \\
 & & - \Bra \De^{\star} \Ord(z_1,n_1)\, \De^{\star} \Ord(z_2,n_2) \Ket_{\star} \Bra \De^{\star} \Ord(z_3,n_3) \, \De^{\star} \Ord(z_4,n_4) \Ket_{\star} \nn \\
 & & - \Bra \De^{\star} \Ord(z_1,n_1)\, \De^{\star} \Ord(z_3,n_3) \Ket_{\star} \Bra \De^{\star} \Ord(z_2,n_2) \, \De^{\star} \Ord(z_4,n_4) \Ket_{\star} \nn \\
 & & - \Bra \De^{\star} \Ord(z_1,n_1)\, \De^{\star} \Ord(z_4,n_4) \Ket_{\star} \Bra \De^{\star} \Ord(z_2,n_2) \, \De^{\star} \Ord(z_3,n_3) \Ket_{\star} \, , \nn
\eea
where here the star is a placeholder for ``ob" and ``th", the corresponding averages are respectively $\bra \dots \ket_{\rm SO(3)}$ and $\bra \dots \ket$ and we have omitted the $\vec{x}_o$ dependencies in the ``ob" case for notational simplicity. Clearly, the these functions are invariant under a common rotation of their arguments, so we can compute their harmonic components by projecting on the basis \eqref{eq:GenYbasis}. Using the connected analogue of \eqref{eq:GobofGobl} and also \eqref{eq:Pldef}, \eqref{eq:3jsum1}, \eqref{eq:3jsum2} and the $3-j$ symbol symmetries, we recover the known expressions for the reduced power spectrum, bispecrum and trispectrum estimators \cite{HU01a}\footnote{Note that \eqref{eq:Tob} agrees with Eqs. (26) and (20) of \cite{HU01a}, as the normalization convention is different by an overall factor of $(-1)^L\sqrt{2L+1}$.}
\bea
 & & C^{\rm ob}_l \( \vec{x}_o; z_1, z_2 \) \equiv \frac{\De^{\rm ob} \Ord_{lm}(\vec{x}_o; z_1)\, \De^{\rm ob} \Ord^*_{lm}(\vec{x}_o; z_2)}{2l + 1} \label{eq:Cob} \, , \\
 & & B^{\rm ob}_{l_1l_2l_3}(\vec{x}_o;z_1, z_2, z_3) \equiv \( \begin{array}{ccc} l_1 & l_2 & l_3 \\ m_1 & m_2 & m_3 \end{array} \) \prod_{k=1}^3 \De^{\rm ob} \Ord_{l_k m_k}(\vec{x}_o;z_k) \, ,  \label{eq:Bob} \\
 & & T^{\rm ob}_{l_1l_2l_3l_4|L}(\vec{x}_o;z_1, z_2, z_3,z_4) \nn \\
 & \equiv & (-1)^{L+M} \sqrt{2L+1} \( \begin{array}{ccc} l_1 & l_2 & L \\ m_1 & m_2 & -M \end{array} \) \( \begin{array}{ccc} L & l_3 & l_4 \\ M & m_3 & m_4 \end{array} \) \prod_{k=1}^4 \De^{\rm ob} \Ord_{l_k m_k}(\vec{x}_o;z_k) \nn \\
 & & - \, (-1)^{l_1+l_3} \sqrt{\( 2l_1 + 1 \) \( 2l_3 + 1 \)}\, \de_{l_1l_2} \de_{l_3l_4} \de_{L0} C^{\rm ob}_{l_1}(\vec{x}_o;z_1,z_2)\, C^{\rm ob}_{l_3}(\vec{x}_o;z_3,z_4) \nn \\
 & & - \, (-1)^{l_1 + l_2} \sqrt{2L+1} \, \{  l_1 \,\,\, l_2 \,\,\, L \}\, \de_{l_1l_3} \de_{l_2l_4} C^{\rm ob}_{l_1}(\vec{x}_o;z_1,z_3)\, C^{\rm ob}_{l_2}(\vec{x}_o;z_2,z_4) \nn \\
 & & - \, (-1)^L \sqrt{2L+1} \, \{  l_1 \,\,\, l_2 \,\,\, L \}\,\de_{l_1l_4} \de_{l_2l_3} C^{\rm ob}_{l_1}(\vec{x}_o;z_1,z_4)\, C^{\rm ob}_{l_2}(\vec{x}_o;z_2,z_3)  \, . \label{eq:Tob}
\eea
The theoretical counterparts are then obtained by replacing ``ob" $\to$ ``th" and ensemble averaging, again in accordance with the $N = 2,3,4$ results of \cite{HU01a}.

Let us now see how the linear ergodic relation \eqref{eq:ergh} looks like in terms of the connected parts. First, because of \eqref{eq:obiden}, the observational spectra vanish identically whenever at least one of their $l_k$ entries is zero, i.e. they have vanishing ``monopoles" by construction. In contrast, this is not the case for the theoretical spectra \eqref{eq:thiden}. As a concrete example, let us consider the relation between the connected 2-point correlation functions, by expressing them in terms of the full functions and using \eqref{eq:erg}
\bea
C^{\rm th}(z_1, z_2, n_1, n_2) & \equiv & G^{\rm th}(z_1, z_2, n_1, n_2) - G^{\rm th}(z_1)\, G^{\rm th}(z_2) \nn \\
 & \os{\rm erg.}{=} & \frac{1}{V} \int \ed^3 x_o\, G^{\rm ob}(\vec{x}_o; z_1, z_2, n_1, n_2) - G^{\rm th}(z_1)\, G^{\rm th}(z_2) \nn \\
 & \equiv & \frac{1}{V} \int \ed^3 x_o \[ C^{\rm ob}(\vec{x}_o; z_1, z_2, n_1, n_2) + G^{\rm ob}(\vec{x}_o; z_1)\, G^{\rm ob}(\vec{x}_o; z_2) \] \nn \\
 & & -\, G^{\rm th}(z_1)\, G^{\rm th}(z_2) \nn \\
 & \os{\rm erg.}{=} & \frac{1}{V} \int \ed^3 x_o \, C^{\rm ob}(\vec{x}_o; z_1, z_2, n_1, n_2) + {\rm Cov}\( z_1; z_2 \) \, , \label{eq:Crealrel}
\eea
where ${\rm Cov}\( z_1; z_2 \)$ is the covariance matrix of the 1-point function with itself (see \eqref{eq:sigeo}). Alternatively, in harmonic space 
\beq \label{eq:Crel}
C^{\rm th}_l(z_1,z_2) \os{\rm erg.}{=} \frac{1}{V} \int \ed^3 x_o \, C^{\rm ob}_l(\vec{x}_o;z_1,z_2) + \de_{l0} {\rm Cov}(z_1; z_2) \, . 
\eeq
The difference with \eqref{eq:ergh} in the connected case is a monopole term, which is also exactly the multipole for which the observational spectrum vanishes identically $C^{\rm ob}_0(z_1,z_2) \equiv 0$. Therefore, a better display of \eqref{eq:Crel} could be
\beq
C^{\rm th}_{l>0}(z_1,z_2) \os{\rm erg.}{=} \frac{1}{V} \int \ed^3 x_o \, C^{\rm ob}_{l>0}(\vec{x}_o;z_1,z_2) \, , \hspace{1cm} C^{\rm th}_0(z_1,z_2) \os{\rm erg.}{=} {\rm Cov}(z_1; z_2) \, .
\eeq 
Thus, the ergodic relation \eqref{eq:ergh} still holds for the components containing physical information $C^{\rm ob}_l \neq 0$, while the left-over equation relates the covariance matrix of the 1-point function to the monopole of the theoretical power spectrum. As a result, the absolute 1-sigma error \eqref{eq:sidef} associated with the 1-point function approximation 
\beq
G^{\rm ob}(x_o; z) \approx G^{\rm th}(z) \pm \Si(z) \, ,
\eeq 
is simply
\beq 
\Si(z) := \sqrt{{\rm Cov}(z; z)} \os{\rm erg.}{=} \sqrt{C^{\rm th}_0(z,z)} \, .
\eeq
This picture generalizes to the case of higher $N$, as one can check by expressing $\De^{\rm th} \Ord_{lm}$ in terms of $\De^{\rm ob} \Ord_{lm}$ in the theoretical spectra $B^{\rm th}$ and $T^{\rm th}$. One finds that the resulting equation splits into two sets of equations: the non-monopole ones, i.e. the ones where none of the $l_k$ is zero and for which \eqref{eq:ergh} still holds, and the ones relating the theoretical monopoles to the covariance matrix and higher order analogues (skewness, kurtosis, etc.) of lower-$N$ spectra. For this reason, from now on all of our equations will hold up to monopole terms, so that we do not have to deal with this subtlety. 

The next non-trivial case is the trispectrum, where one now subtracts products of 2-point functions. For the purposes of our point it suffices to describe it schematically as
\beq
T^{\rm th}\( \{ n_k, z_k \}_{k=1}^4 \) \equiv \Bra \prod_{k=1}^4 \De^{\rm th} \Ord(\vec{x}_o; z_k,n_k) \Ket - C^{\rm th}(z_1,z_2,n_1,n_2)\, C^{\rm th}(z_3,z_4,n_3,n_4) - \dots
\eeq
where the ellipses here denote the other two possible orderings of the 2-point function arguments. In complete analogy with the derivation in the $N = 2$ case \eqref{eq:Crealrel}, using \eqref{eq:erg} we find
\bea
T^{\rm th}\( \{ n_k, z_k \}_{k=1}^4 \) & \equiv & G^{\rm th}\( \{ n_k,z_k \}_{k=1}^4 \) - G^{\rm th}(z_1,z_2,n_1,n_2)\, G^{\rm th}(z_3,z_4,n_3,n_4) - \dots + {\rm mon.} \nn \\
 & \os{\rm erg.}{=} & \frac{1}{V} \int \ed^3 x_o \, G^{\rm ob}\( \vec{x}_o; \{ n_k,z_k \}_{k=1}^4 \)   \nn \\
 & & -\, G^{\rm th}(z_1,z_2,n_1,n_2)\, G^{\rm th}(z_3,z_4,n_3,n_4) - \dots + {\rm mon.} \nn \\
 & \equiv & \frac{1}{V} \int \ed^3 x_o \[ C^{\rm ob}\( \vec{x}_o; \{ n_k,z_k \}_{k=1}^4 \)  \right. \nn \\
 & & \hspace{2cm} \left. +\, G^{\rm ob}(\vec{x}_o;z_1,z_2,n_1,n_2)\, G^{\rm ob}(\vec{x}_o;z_3,z_4,n_3,n_4) + \dots \] \nn \\
 & & - \, G^{\rm th}(z_1,z_2,n_1,n_2)\, G^{\rm th}(z_3,z_4,n_3,n_4) - \dots + {\rm mon.} \nn \\
 & \os{\rm erg.}{=} & \frac{1}{V} \int \ed^3 x_o \, C^{\rm ob}\( \vec{x}_o; \{ n_k,z_k \}_{k=1}^4 \) \nn \\
 & & +\, {\rm Cov}(z_1,z_2,n_1,n_2;z_3,z_4,n_3,n_4) + \dots + {\rm mon.} 
\eea
so the difference is again a covariance matrix, but now the one of the 2-point function with itself. In harmonic space we then find
\bea 
T^{\rm th}_{l_1l_2l_3l_4|L}\( z_1, z_2, z_3, z_4 \) & \os{\rm erg.}{=} & \frac{1}{V} \int \ed^3 x_o \, T^{\rm ob}_{l_1l_2l_3l_4|L}\( \vec{x}_o; z_1, z_2, z_3, z_4 \) \label{eq:erghT} \\
 & & + \, (-1)^{l_1+l_3} \sqrt{\( 2l_1 + 1 \) \( 2l_3 + 1 \)}\, \de_{l_1l_2} \de_{l_3l_4} \de_{L0} {\rm Cov}_{l_1;l_3}(z_1,z_2;z_3,z_4) \nn \\
 & & + \, (-1)^{l_1 + l_2} \sqrt{2L+1} \, \{  l_1 \,\,\, l_2 \,\,\, L \}\, \de_{l_1l_3} \de_{l_2l_4} {\rm Cov}_{l_1;l_2}(z_1,z_3;z_2,z_4) \nn \\
 & & + \, (-1)^L \sqrt{2L+1} \, \{  l_1 \,\,\, l_2 \,\,\, L \}\,\de_{l_1l_4} \de_{l_2l_3} {\rm Cov}_{l_1;l_2}(z_1,z_4;z_2,z_3) + {\rm mon.} \nn
\eea
where we recognize the same structure as in \eqref{eq:Tob} by construction and we have the covariance matrix of the 2-point spectra 
\beq \label{eq:Covllp}
{\rm Cov}_{l;l'}(z_1,z_2;z_3,z_4) \os{\rm erg.}{=} \frac{1}{V} \int \ed^3 x_o \, G_l^{\rm ob}(\vec{x}_o;z_1,z_2)\, G_{l'}^{\rm ob}(\vec{x}_o;z_3,z_4) - G_l^{\rm th}(z_1, z_2) \, G_{l'}^{\rm th}(z_3, z_4)   \, .   
\eeq
A first important difference with the 2-point case \eqref{eq:Crel} is that the extra terms here affect multipoles of arbitrary magnitude, although only a small subset, the one of pairwise equal $l_k$'s. A second important difference is that the components of $T^{\rm ob}$ with pairwise equal $l_k$'s are not identically zero in general, i.e. the equations no longer split as in the monopole case. One must therefore take \eqref{eq:erghT} as it is and infer that the analogue of the single-observer approximation \eqref{eq:stdergh} in this case is actually
\bea
T^{\rm ob}_{l_1l_2l_3l_4|L}\( \vec{x}_o; z_1, z_2, z_3, z_4 \) & \os{\rm erg.}{\approx} & T^{\rm th}_{l_1l_2l_3l_4|L}\( z_1, z_2, z_3, z_4 \) \label{eq:singobs4p} \\
 & & - \, (-1)^{l_1+l_3} \sqrt{\( 2l_1 + 1 \) \( 2l_3 + 1 \)}\, \de_{l_1l_2} \de_{l_3l_4} \de_{L0} {\rm Cov}_{l_1;l_3}(z_1,z_2;z_3,z_4) \nn \\
 & & - \, (-1)^{l_1 + l_2} \sqrt{2L+1} \, \{  l_1 \,\,\, l_2 \,\,\, L \}\, \de_{l_1l_3} \de_{l_2l_4} {\rm Cov}_{l_1;l_2}(z_1,z_3;z_2,z_4) \nn \\
 & & - \, (-1)^L \sqrt{2L+1} \, \{  l_1 \,\,\, l_2 \,\,\, L \}\,\de_{l_1l_4} \de_{l_2l_3} {\rm Cov}_{l_1;l_2}(z_1,z_4;z_2,z_3) + {\rm mon.} \nn
\eea 
Finally, just as the 2-point case \eqref{eq:Crel} with $l=0$ provides the cosmic variance of the 1-point function for free, the 4-point case \eqref{eq:erghT} with $l_1 = l_2 =: l$, $l_3 = l_4 =: l'$ and $L=0$ provides the cosmic variance for the power spectrum. However, here $T^{\rm ob}_{lll'l'|0} \neq 0$, but rather (use \eqref{eq:Tob} and properties of the $3-j$ symbols)
\beq
T^{\rm ob}_{lll'l'|0}(\vec{x}_o;z_1, z_2, z_3,z_4) \equiv - \de_{ll'} \[ C^{\rm ob}_l \( \vec{x}_o; z_1,z_3 \) C^{\rm ob}_l \( \vec{x}_o; z_2,z_4 \) + \( z_3 \leftrightarrow z_4 \) \] \, ,
\eeq
so, inserting this in \eqref{eq:erghT}, using \eqref{eq:singobs4p} and isolating the covariance matrix, we find
\bea
{\rm Cov}_{l;l'}(z_1,z_2,z_3,z_4) & \equiv & \frac{(-1)^{l+l'}}{\sqrt{(2 l + 1)\, (2l' + 1)}} \[ \frac{}{} \de_{ll'} \[ C_l^{\rm th}(z_1,z_3)\, C_l^{\rm th}(z_2,z_4) + \( z_3 \leftrightarrow z_4 \) \] \right. \nn \\
 & & \left. \hspace{3.5cm}  +\, T^{\rm th}_{lll'l'|0}(z_1,z_2,z_3,z_4) \frac{}{} \] + {\rm mon.}  \label{eq:Covllp}
\eea
This generalizes the result of \cite{HU01a}, worked out for the CMB where there is no redshift dependence, to the case of generic redshifts along the light-cone. In particular, the cosmic variance of the power spectrum, i.e. the absolute 1-sigma error \eqref{eq:sidef} in the approximation
\beq
C_l^{\rm ob}(\vec{x}_o;z,z') \approx C_l^{\rm th}(z,z') \pm \Si_l(z,z') \, ,
\eeq
is
\bea
\Si^2_l(z,z') & := & {\rm Cov}_{l;l}(z,z',z,z')  \label{eq:Si2ps} \\
 & \equiv & \frac{1}{2l+1} \[ \[ C_l^{\rm th}(z,z') \]^2 + C_l^{\rm th}(z,z)\, C_l^{\rm th}(z',z') + T^{\rm th}_{llll|0}(z,z',z,z') \] + {\rm mon.}   \nn
\eea
Finally, for Gaussian statistics and equal redshifts, we recover the well-known result\footnote{See e.g. \cite{Durrer:2008eom} for the case of the CMB power spectrum.}
\beq \label{eq:SiGausszz}
\Si_l(z,z)|_{\rm Gauss.} = \sqrt{\frac{2}{2l+1}}\, C_l^{\rm th}(z,z)  \, .
\eeq
Given the discussion of subsection \ref{sec:clth} and comparing \eqref{eq:Si2ps} with \eqref{eq:Silgen}, we find that here the central limit theorem essentially states that the trispectrum decays faster with growing $l$ than the $C_l^2$ terms.

\subsection{Mind the monopole when using relative fluctuations} \label{sec:mono}

Until now we have considered the ``absolute" fluctuations $\De^{\rm ob} \Ord$ and $\De^{\rm th} \Ord$, but in practice the most convenient ones to work with are the relative ones (if $G^{\rm th}(z) \neq 0$) 
\beq
\de^{\rm ob} \Ord(\vec{x}_o;z,n) := \frac{\De^{\rm ob} \Ord(\vec{x}_o;z,n)}{G^{\rm ob}(\vec{x}_o;z)} \, , \hspace{1cm}  \de^{\rm th} \Ord(\vec{x}_o;z,n) := \frac{\De^{\rm th} \Ord(\vec{x}_o;z,n)}{G^{\rm th}(z)} \, .
\eeq
This introduces a non-linear difference between the observational and theoretical definitions and  will  affect the covariance matrices of the corresponding spectra. Indeed, consider for instance the case of the relative power spectra
\bea
\ti{C}^{\rm ob}_l(\vec{x}_o; z_1,z_2) & := & \frac{\de^{\rm ob} \Ord_{lm}(\vec{x}_o;z_1)\, \de^{\rm ob} \Ord^*_{lm}(\vec{x}_o;z_2)}{2l+1} \equiv \frac{C^{\rm ob}_l(\vec{x}_o;z_1,z_2)}{G^{\rm ob}(\vec{x}_o; z_1)\, G^{\rm ob}(\vec{x}_o; z_2)} \, , \\
\ti{C}^{\rm th}_l(z_1,z_2) & \os{\rm erg.}{=} & \frac{1}{V} \int \ed^3 x_o \,\frac{\de^{\rm th} \Ord_{lm}(\vec{x}_o;z_1)\, \de^{\rm th} \Ord^*_{lm}(\vec{x}_o;z_2)}{2l+1} \equiv  \frac{C^{\rm th}_l(z_1,z_2)}{G^{\rm th}(z_1)\, G^{\rm th}(z_2)} \, ,
\eea
which are denoted using tilded letters to distinguish them from the absolute ones $C_l^{\rm ob}$ and $C_l^{\rm th}$. The corresponding covariance matrix is given by
\beq
\widetilde{\rm Cov}_{l;l'}(z_1,z_2,z_3,z_4) \os{\rm erg.}{=} \frac{1}{V} \int \ed^3 x_o \[ \ti{C}_l^{\rm ob}(\vec{x}_o;z_1,z_2) - \ti{C}_l^{\rm th}(z_1, z_2) \] \[ \ti{C}_{l'}^{\rm ob}(\vec{x}_o;z_3,z_4) - \ti{C}_{l'}^{\rm th}(z_3, z_4) \] \, ,
\eeq
and one would naively expect this to be simply
\beq \label{eq:SitSiwrongrel}
\widetilde{\rm Cov}_{l;l'}(z_1,z_2,z_3,z_4) = \frac{{\rm Cov}_{l;l'}(z_1,z_2,z_3,z_4)}{G^{\rm th}(z_1)\, G^{\rm th}(z_2)\, G^{\rm th}(z_3)\, G^{\rm th}(z_4)} + {\rm mon.} \, ,
\eeq
where here ${\rm Cov}_{l;l'}$ is the covariance matrix of the absolute 2-point spectrum \eqref{eq:Covllp}. In particular, one would then infer the analogue of \eqref{eq:SiGausszz} for the corresponding cosmic variance in the equal redshift Gaussian case
\beq  \label{eq:tSiGausszz}
\ti{\Si}_l(z,z)|_{\rm Gauss.} = \sqrt{\frac{2}{2l+1}}\, \ti{C}_l^{\rm th}(z,z) \, .
\eeq
However, as we will now show, Eqs. \eqref{eq:SitSiwrongrel} and \eqref{eq:tSiGausszz} are actually only approximate, because in order to obtain them one must wrongly assume that $G^{\rm ob}(\vec{x}_o;z) = G^{\rm th}(z)$ or, according to \eqref{eq:obiden}, that the monopole of the observable is zero at $\vec{x}_o$ for all $z$
\beq
\de^{\rm th} \Ord_{00}(\vec{x}_o, z) = 0 \, . 
\eeq
Indeed, 
\bea
 & & {\rm Cov}_{l;l'}(z_1,z_2,z_3,z_4) \\
 & \os{\rm erg.}{=} & \frac{1}{V} \int \ed^3 x_o \[ C_l^{\rm ob}(\vec{x}_o;z_1,z_2)\, C_{l'}^{\rm ob}(\vec{x}_o;z_3,z_4) \] - C^{\rm th}_l(z_1,z_2)\, C^{\rm th}_{l'}(z_3,z_4) + {\rm mon.} \\
 & \equiv & \frac{1}{V} \int \ed^3 x_o\, G^{\rm ob}(\vec{x}_o;z_1)\, G^{\rm ob}(\vec{x}_o;z_2)\, G^{\rm ob}(\vec{x}_o;z_3)\, G^{\rm ob}(\vec{x}_o;z_4) \, \ti{C}_l^{\rm ob}(\vec{x}_o;z_1,z_2)\, \ti{C}_{l'}^{\rm ob}(\vec{x}_o;z_3,z_4) \nn \\
 & &
 - \, G^{\rm th}(z_1)\, G^{\rm th}(z_2)\, G^{\rm th}(z_3)\, G^{\rm th}(z_4) \, \ti{C}^{\rm th}_l(z_1,z_2) \, \ti{C}^{\rm th}_{l'}(z_3,z_4) + {\rm mon.}  \nn \\
 & \equiv & G^{\rm th}(z_1)\, G^{\rm th}(z_2) \, G^{\rm th}(z_3)\, G^{\rm th}(z_4) \[ \widetilde{\rm Cov}_{ll'}(z_1,z_2,z_3,z_4) + \ti{R}_{ll'}(z_1,z_2,z_3,z_4) \]  + {\rm mon.} \, , \nn
\eea
where the remainder term $\ti{R}_{ll'}$ depends on the monopole $\de^{\rm th} \Ord_{00}(\vec{x}_o;z_k)$ that comes from converting the $G^{\rm ob}(\vec{x}_o,z_k)$ into $G^{\rm ob}(z_k)$ and which cannot go through the integral over $\vec{x}_o$, because of its dependence on that variable. The remainder term is therefore controlled by a combination of monopoles of the theoretical power spectrum, bispectrum and trispectrum which, unlike the extra ``mon." terms described in the previous section, contribute here to {\it all} $l$ values, not just to $l = 0$. Thus, it is important to know whether one compares absolute quantities or relative ones, because the corresponding covariance matrices are different and therefore so will be the results of the corresponding Fisher forecasts.

\section{Conclusion} \label{sec:discussion}

In this paper we have addressed two potentially important issues for the next stages of precision cosmology. First, we have developed the general theory of the rotationally-invariant (``reduced") angular $N$-point spectra. The method employed so far \cite{HU01a} is not suited for obtaining expressions for arbitrary $N$, because it requires solving some increasingly complicated algebraic equation as $N$ grows. As a result, one must work out each $N$ case separately and the present literature contains a detailed description only up to the $N = 4$ case \cite{HU01a}. We have presented an alternative construction which provides all the relevant definitions and relations associated with reduced angular $N$-point spectra straightforwardly and for all $N$. This includes the covariance matrix of these spectra for arbitrary $N$ and $N'$, which is controlled by a squeezed $(N + N')$-point spectrum and leads to a cosmic variance of the $N$-point spectrum that decays as $l^{(N-1)/2}$ for large $l_k$ numbers. Our construction is based on the introduction of the ``multilateral'' Wigner symbols, which generalize the Wigner $3-j$ symbols of triangles to polygons. With these we have built an orthonormal harmonic basis for $N$-point correlation functions on the sphere, such that the reduced spectra appear as the coefficients in this basis. We have also discussed a corresponding diagrammatic representation, generalizing the one of \cite{HU01a,ABPE10} for the cases $N=4,5$, which was also considered in \cite{LACAS14} in the flat sky limit. Even though the determination of a generic $N$-point spectrum might be prohibitive numerically, our general framework allows easily for specialization, e.g. to squeezed spectra with only one $l$ varying while the others are fixed to some low value. 

The second important part of the paper consists in motivating and justifying the consideration of observer terms of cosmological observables inside the ensemble averages of correlation functions and spectra. Through a careful derivation of the relation between theoretical predictions (``$G^{\rm th}$") and observations (``$G^{\rm ob}$"), under the assumptions of statistical homogeneity and isotropy and ergodicity, we have shown that no special treatment of the observer point is required whatsoever on the theoretical side. Ensemble averaging field products at this point does not introduce any inconsistencies, nor does it imply any new assumptions or uncertainties. This motivates us to consider the {\it full} analytical expressions of cosmological observables in calculations, especially at non-linear order in perturbations where observer terms are intertwined with source and line-of-sight contributions in a non-trivial way, as we argued in subsection \ref{sec:obsissue}. Thus, with this conceptual ambiguity being lifted, we can now safely state that a rigorous treatment of cosmological observables should include the observer terms, or at least a check of the magnitude of their contribution in the final result if they are neglected.  

We have also taken advantage of the present framework to discuss some subtle aspects of working with the connected part of spectra, as one usually does, when relating theory to observation. When working directly with fluctuations of observables around their average (1-point function) value, as is often the case in cosmology, the difference between full and connected spectra is simply a sum of products of spectra with $N > 1$. However, here we pointed out that the notion of fluctuation is different in the theoretical and observational cases, since the 1-point functions are different (ensemble versus sky average). As a result, the relation between theoretical and observational {\it connected} spectra involves extra ``monopole" terms. Moreover, if one works with relative, instead of absolute, fluctuations of observables, then one must take into account additional corrections which now affect all multipoles, not just the monopoles. A direct consequence of this fact is that the covariance matrices computed with the relative and absolute fluctuations are not proportional to each other, but have an additional remainder term. Thus, this can lead to different results in the corresponding Fisher analyses, depending on whether one uses absolute or relative observational fluctuations. 

Finally, and remarkably, the fact that the observational and theoretical 1-point functions are not equal in general might also be relevant in the context of parameter estimation involving CMB data. Indeed, the standard treatment of the CMB temperature is to simply neglect cosmic variance and therefore consider these two quantities as equal $\bar{T} := \Bra T \Ket = \Bra T \Ket_{\rm SO(3)}(\vec{x}_o)$. This is justified by the high precision of the average CMB temperature measurement \cite{FIRAS94} and, most importantly, by the smallness of the associated relative cosmic variance $C_0 \sim 10^{-5}$, which sets the fundamental uncertainty floor. In principle, however, one should consider $\bar{T}$ (or $\om_{\ga}$) as an extra cosmological parameter to be varied in the likelihood analysis. Through a careful Fisher analysis that incorporates the above-mentioned subtleties \cite{Yoo:2019qsl}, the impact of an independent $\bar{T}$ on the other cosmological parameters and their errors is estimated in \cite{Yoo:2019dyl}.

\acknowledgments

We are grateful to Fabien Lacasa and Thiago Pereira for pointing out references \cite{LACAS14} and \cite{ABPE10}, respectively, and for useful comments. We acknowledge support by the Swiss National Science Foundation. J.Y. and E.M. are further supported by a Consolidator Grant of the European Research Council (ERC-2015-CoG grant 680886).

\appendix

\section{Harmonic decomposition of ${\rm SO(3)}$-invariant functions on $S_2^N$}  \label{sec:hdec}

Consider a function $f$ on $S_2^N$ that satisfies
\beq \label{eq:finvar}
f(n_1, \dots, n_N) \equiv f(R n_1, \dots, R n_N)  \, ,
\eeq
for arbitrary rotations $R$. We can then average over the SO(3) group on the right-hand side (see \eqref{eq:SO3avdef})
\beq \label{eq:feqfso3}
f(n_1, \dots, n_N) \equiv \bra f(n_1, \dots, n_N) \ket_{\rm SO(3)}  \, .
\eeq
We then decompose each one of the $n_k$ dependencies into spherical harmonics
\beq
f(n_1, \dots, n_N) \equiv f^{l_1 \dots l_N}_{m_1 \dots m_N} Y_{l_1 m_1}(n_1) \dots Y_{l_N m_N}(n_N) \, , 
\eeq
where
\beq
f^{l_1 \dots l_N}_{m_1 \dots m_N} := \int \( \prod_{k=1}^N \frac{\ed \Om_k}{4\pi} \, Y_{l_k m_k}^*(n_k) \) f(n_1, \dots, n_N) \, .
\eeq
The summation over $l,m$ indices will be kept implicit for notational simplicity. In what follows, we will encounter both dummy and free $l,m$ indices, so their nature will be inferable by looking at both sides of the equation. The $m$ indices will always be clearly associated to some $l$ value and therefore run from $-l$ to $l$, while the $l$ indices run from $0$ to $\infty$. Also, note that we use the less conventional normalization of spherical harmonics
\beq \label{eq:Ynorm}
\frac{1}{4\pi} \int \ed \Om\, Y_{lm}(n)\, Y^*_{l'm'}(n) \equiv \de_{ll'} \de_{mm'}  \, .
\eeq
Equation \eqref{eq:feqfso3} now reads
\beq \label{eq:Ishdec}
f(n_1, \dots, n_N) \equiv f^{l_1 \dots l_N}_{m_1 \dots m_N} \times \frac{1}{8\pi^2} \int \ed R  \prod_{k=1}^N Y_{l_k m_k}(R^{-1} n_k)  \, , 
\eeq
where the Haar measure $\ed R$ is given in \eqref{eq:HaarSO3}. We can then extract the $R$-dependence out of the spherical harmonics by using their transformation property under rotations
\beq \label{eq:Ytrans}
Y_{lm}\( R^{-1}(\al,\be,\ga)\, n \) \equiv Y_{lm'} ( n ) \, D_{l,m'm}(\al,\be,\ga) \, ,
\eeq 
where the $D_l$ are the Wigner matrices forming the $(2l+1)$-dimensional irreducible representation of SO(3)
\beq
D_l(R)\, D_l(R') \equiv D_l(R R') \, , \hspace{1cm} D_l(R)\, D^{\dagger}_l(R) \equiv \mathbb{I} \, .
\eeq
We thus have
\beq \label{eq:GobofI}
f(n_1, \dots, n_N) \equiv I^{l_1 \dots l_N}_{m'_1 \dots m'_N, m_1 \dots m_N} f^{l_1 \dots l_N}_{m_1 \dots m_N} \prod_{k=1}^N Y_{l_k m'_k}(n_k) \, , 
\eeq
where 
\beq \label{eq:IDlprod}
I^{l_1 \dots l_N}_{m'_1 \dots m'_N, m_1 \dots m_N} := \frac{1}{8\pi^2} \int_0^{2\pi} \ed \al \int_0^{\pi} \sin \be\, \ed \be \int_0^{2\pi} \ed \ga \prod_{k=1}^N D_{l_k, m'_k m_k}(\al,\be,\ga) \, .
\eeq
In the $N=2$ case, we can use the identity
\beq \label{eq:Dlstar}
D^*_{l,mm'} \equiv (-1)^{m+m'} D_{l,-m-m'} \, , 
\eeq
and the orthonormality relation
\beq \label{eq:Dortho}
\frac{1}{8\pi^2} \int_0^{2\pi} \ed \al \int_0^{\pi} \sin \be\, \ed \be \int_0^{2\pi} \ed \ga \, D_{l_1, m_1 m'_1}(\al,\be,\ga)\, D^*_{l_2,m_2m'_2}(\al,\be,\ga) \equiv \frac{1}{2l_1+1}\, \de_{l_1l_2} \de_{m_1 m_2} \de_{m'_1 m'_2}  \, ,
\eeq
to obtain
\beq
I^{l_1 l_2}_{m'_1 m'_2, m_1 m_2} \equiv \frac{(-1)^{m_1 + m'_1}}{2l_1+1}\, \de_{l_1l_2} \de_{-m_1 m_2} \de_{-m'_1 m'_2} \, ,
\eeq
and thus
\beq
f(n_1, n_2) \equiv f_l \( 2 l +1 \) P_l \( n_1 \cdot n_2 \)
\eeq
where the $P_l$ are the Legendre polynomials, arising from the identity
\beq \label{eq:Pldef}
P_l \( n_1 \cdot n_2 \) \equiv \frac{1}{2l+1} Y_{lm}(n_1)\, Y^*_{lm}(n_2) \, ,
\eeq
and
\beq 
f_l := \frac{(-1)^m f^{ll}_{m,-m}}{2l + 1} \, .
\eeq
In the case of observable products, we obtain the well-known results \eqref{eq:GobofGobl} and \eqref{eq:Gob2l}. For the case of most interest $N > 2$, we need to use iteratively the ``Clebsch-Gordan" composition rule
\beq \label{eq:CG}
D_{l_1, m_1 m'_1} D_{l_2, m_2 m'_2} = \( \begin{array}{ccc} l_1 & l_2 & L \\ m_1 & m_2 & -M \end{array} \)  \( \begin{array}{ccc} l_1 & l_2 & L \\ m'_1 & m'_2 & -M' \end{array} \) \( 2 L +1 \) (-1)^{M+M'} D_{L, M M'} \, ,
\eeq
in order to reduce the product in \eqref{eq:IDlprod} down to a single pair, in which case we can use again \eqref{eq:Dlstar} and \eqref{eq:Dortho}. The result is
\beq \label{eq:IofW2}
I^{l_1 \dots l_N}_{m'_1 \dots m'_N, m_1 \dots m_N} \equiv W^{l_1 \dots l_N|L_1 \dots L_{N-3}}_{m'_1 \dots m'_N} W^{l_1 \dots l_N|L_1 \dots L_{N-3}}_{m_1 \dots m_N} \, ,
\eeq
where the coefficients on the right-hand side are given in \eqref{eq:Wdef}. Going back to \eqref{eq:GobofI}, this finally implies the decomposition
\beq \label{eq:ffdecomp}
f(n_1, \dots, n_N) \equiv f_{l_1 \dots l_N| L_1 \dots L_{N-3}} Y_{l_1 \dots l_N| L_1 \dots L_{N-3}} \( n_1, \dots, n_N \)  \, , 
\eeq
where
\beq  
f_{l_1 \dots l_N| L_1 \dots L_{N-3}} := W^{l_1 \dots l_N| L_1 \dots L_{N-3}}_{m_1 \dots m_N} f^{l_1 \dots l_N}_{m_1 \dots m_N} \, ,
\eeq
and
\beq
Y_{l_1 \dots l_N| L_1 \dots L_{N-3}} \( n_1, \dots, n_N \) := W^{l_1 \dots l_N| L_1 \dots L_{N-3}}_{m_1 \dots m_N} \prod_{k=1}^N Y_{l_k m_k}(n_k)  \, . 
\eeq
The latter form a basis for SO(3)-invariant functions on $S_2^N$. Indeed, we have just shown that they generate that space, so we must still show that they are orthogonal under the natural scalar product. Using the identity \eqref{eq:3jsum1}, where
\beq \label{eq:Tdelta1}
\{ l_1 \,\,\, l_2 \,\,\, l_3 \} := \left\{ \begin{array}{cc} 1 & \text{if  } l_1 \in \{ |l_2 - l_3|, \dots, l_2 + l_3 \} \\ 0 & \text{otherwise} \end{array} \right. \, ,
\eeq
one obtains
\beq \label{eq:Wortho}
W^{l_1 \dots l_N| L_1 \dots L_{N-3}}_{m_1 \dots m_N} W^{l_1 \dots l_N| L'_1 \dots L'_{N-3}}_{m_1 \dots m_N} \equiv \{ l_1 \dots l_N| L_1 \dots L_{N-3} \} \prod_{k=1}^{N-3} \de^{L_k L'_k}  \, ,
\eeq
where we have defined
\beq \label{eq:Mdelta}
\{ l_1 \dots l_N| L_1 \dots L_{N-3} \} := \{ l_1 \,\,\, l_2 \,\,\, L_1 \} \( \prod_{k=1}^{N-4} \{ L_k \,\,\, l_{k+2} \,\,\, L_{k+1} \} \) \{ L_{N-3} \,\,\, l_{N-1} \,\,\, l_N \} \, ,
\eeq
and therefore the orthonormality relation
\bea 
 & & \int \( \prod_{k=1}^N \frac{\ed \Om_k}{4\pi} \) Y_{l_1 \dots l_N| L_1 \dots L_{N-3}}\( n_1, \dots, n_N \) Y^*_{l'_1 \dots l'_N| L'_1 \dots L'_{N-3}}\( n_1, \dots, n_N \) \nn \\
 & \equiv & \{ l_1 \dots l_N| L_1 \dots L_{N-3} \} \prod_{k=1}^N \de^{l_k l'_k} \prod_{k=1}^{N-3} \de^{L_k L'_k}   \, . \label{eq:newYbortho}
\eea 
Finally, one can use this to invert \eqref{eq:ffdecomp}
\beq \label{eq:fhoff}
f_{l_1 \dots l_N| L_1 \dots L_{N-3}} \equiv \int \(  \prod_{k=1}^N \frac{\ed \Om_k}{4\pi} \) Y^*_{l_1 \dots l_N| L_1 \dots L_{N-3}}\( n_1, \dots, n_N \) f \( n_1, \dots, n_N \) \, .
\eeq

\section{Some  identities for Wigner $3-j$ symbols} \label{sec:3jiden}

Here we display all the identities satisfied by the Wigner $3-j$ symbols that are used in this paper. 

\begin{itemize}

\item
Special cases
\beq \label{eq:3j0}
\( \begin{array}{ccc} l_1 & 0 & 0 \\ m_1 & 0 & 0 \end{array} \) \equiv \de^{l_1}_0 \de^0_{m_1} \, , \hspace{1cm} \( \begin{array}{ccc} l_1 & l_2 & 0 \\ m_1 & m_2 & 0 \end{array} \) \equiv \frac{(-1)^{l_1+m_1}}{\sqrt{2 l_1 + 1}}\, \de^{l_1l_2} \de_{m_1, -m_2} \, .
\eeq

\item
Column exchange symmetry:

even permutations
\beq \label{eq:3jperme}
\( \begin{array}{ccc} l_1 & l_2 & l_3 \\ m_1 & m_2 & m_3 \end{array} \) \equiv \( \begin{array}{ccc} l_2 & l_3 & l_1 \\ m_2 & m_3 & m_1 \end{array} \) \equiv \( \begin{array}{ccc} l_3 & l_1 & l_2 \\ m_3 & m_1 & m_2 \end{array} \) \, ,
\eeq
odd permutations
\beq \label{eq:3jpermo}
\( \begin{array}{ccc} l_1 & l_2 & l_3 \\ m_1 & m_2 & m_3 \end{array} \) \equiv (-1)^{l_1+l_2+l_3} \( \begin{array}{ccc} l_2 & l_1 & l_3 \\ m_2 & m_1 & m_3 \end{array} \) \equiv (-1)^{l_1+l_2+l_3} \( \begin{array}{ccc} l_1 & l_3 & l_2 \\ m_1 & m_3 & m_2 \end{array} \) \, ,
\eeq

\item
$m$-sign flip symmetry
\beq \label{eq:3jmsf}
\( \begin{array}{ccc} l_1 & l_2 & l_3 \\ -m_1 & -m_2 & -m_3 \end{array} \) \equiv (-1)^{l_1+l_2+l_3} \( \begin{array}{ccc} l_1 & l_2 & l_3 \\ m_1 & m_2 & m_3 \end{array} \)  \, ,
\eeq

\item
Sum identities
\beq \label{eq:3jsum1}
\( \begin{array}{ccc} l_1 & l_2 & l_3 \\ m_1 & m_2 & m_3 \end{array} \) \( \begin{array}{ccc} l_1 & l_2 & l'_3 \\ m_1 & m_2 & m'_3 \end{array} \) \equiv \frac{1}{2l_3+1}\, \{ l_1 \,\,\, l_2 \,\,\, l_3 \}\, \de_{l_3l'_3} \de_{m_3 m'_3} \, ,
\eeq
\beq \label{eq:3jsum2}
(-1)^m \( \begin{array}{ccc} L & l & l \\ M & m & -m \end{array} \) \equiv \sqrt{2 l + 1} \, (-1)^l \de_0^L \de_M^0 \, .
\eeq

\end{itemize}

\bibliographystyle{JHEP}
\bibliography{Main}

\end{document}